\documentclass[twocolumn,floatfix,pra,showpacs,preprintnumbers,superscriptaddress,amsmath,amssymb,10pt,aps]{revtex4}
\usepackage{graphicx} 
\usepackage{appendix}
\usepackage{natbib}
\usepackage{amsmath}
\usepackage{bbm}
\usepackage{hyperref}
\usepackage{xurl}
\hypersetup{breaklinks=true}
\usepackage{float}

\begin{document}
\title{Analytical determination of multi-time correlation functions in quantum chaotic systems}  
\def\correspondingauthor{\footnote{Corresponding author: --}}
\author{Yoana R. Chorbadzhiyska}
\affiliation{Center for Quantum Technologies, Faculty of Physics, Sofia University "St. Kliment Ohridski", 5 James Bourchier Blvd, Sofia 1164, Bulgaria}
\author{Peter A. Ivanov}
\affiliation{Center for Quantum Technologies, Faculty of Physics, Sofia University "St. Kliment Ohridski", 5 James Bourchier Blvd, Sofia 1164, Bulgaria}
\author{Charlie Nation}
\affiliation{Department of Physics and Astronomy, University of Exeter, Stocker Road, Exeter EX4 4QL, United Kingdom}

\begin{abstract}
The time-dependence of multi-point observable correlation functions are essential quantities in analysis and simulation of quantum dynamics. Open quantum systems approaches utilize two-point correlations to describe the influence of an environment on a system of interest, and in studies of chaotic quantum system, the out-of-time-ordered correlator (OTOC) is used to probe chaoticity of dynamics. In this work we analytically derive the time dependence of multi-point observable correlation functions in quantum systems from a random matrix theoretic approach, with the highest order function of interest being the OTOC. We find in each case that dynamical contributions are related to a simple function, related to the Fourier transform of coarse-grained wave-functions. We compare the predicted dynamics to exact numerical experiments in a spin chain for various physical observables. We comment on implications towards the emergence of Markovianity and quantum regression in closed quantum systems, as well as relate our results to known bounds on chaotic dynamics.

\end{abstract}

\maketitle

\section{Introduction} 

Since the development of quantum theory there has been debate around the fundamental question of how evolution to thermal equilibrium arises in quantum systems evolving under unitary evolution \cite{schrodinger, von_neumann}. Typical approaches to justifying thermodynamic behavior rely on coupling of a system to a larger environment at thermal equilibrium, however the ability of modern experiments to observe thermalisation in single realizations of closed systems, initialised in pure states \cite{clos_time-resolved_2016, So_Trapped_ion, prufer_condensation_2022, fischer_dynamical_2024,gring,kaufman}, imply a deeper mechanism for thermalisation. Interest in the foundations of quantum statistical physics has thus been rekindled by experimental platforms able to observe quantum dynamics in ever larger systems, and the question of how and when a system may be expected to behave according to thermodynamic rules has become a topic of considerable interest \cite{dalessio_quantum_2016, merali_new_2017, iyoda_fluctuation_2017, deutsch_eigenstate_2018, pappalardi_eigenstate_2022, JoshiProbing2022, dowling_equilibration_2023, strasberg_classicality_2023, munoz-arias_quantum_2024, scandi_thermalisation_2025}.

The Eigenstate Thermalisation Hypothesis (ETH) \cite{deutsch_quantum_1991, srednicki_chaos_1994, rigol2009prl} is understood as a leading mechanism for thermalisation in closed chaotic quantum systems. The ETH can be written as a conjecture on the matrix elements of local observables in the Hamiltonian eigenbasis $H|\psi_\mu\rangle = E_\mu|\psi_\mu\rangle$, as
\begin{equation}
    O_{\mu\nu} = O(E)\delta_{\mu\nu} + \frac{1}{\sqrt{D(E)}} f(E, \omega)\mathcal{R}_{\mu\nu},
\end{equation}
with $O(E)$ being equal to a relevant thermal ensemble at energy $E = \frac{E_\mu + E_\nu}{2}$, $f(E, \omega)$ being a smooth function of energy and the energy gap $\omega = E_\mu - E_\nu$,
$D(E)$ is the density of states, and $\mathcal{R}_{\mu\nu}$ is a random variable with zero mean and unit variance. In words, the ETH can be stated roughly as `eigenstates act as thermal states': a statement that extends beyond simply expectation values guaranteed by the diagonal term, to fluctuations, and even effective temperatures of single eigenstates \cite{Borgonovi_temperature_2017, Nation_snapshots_2020}.
The ETH has since been confirmed in a wide range of non-integrable systems \cite{rigol2009prl, Nation2018, brenes_eigenstate_2020, ivanov2024}. The ETH itself can be derived from weaker assumptions yielding a treatment of a system in terms of `chaotic eigenstates' \cite{Nation2018}, which is the approach taken in this work.

Multi-time observable correlation functions play a vital role in the study of thermalisation processes and the ETH \cite{Bartsch2007OccurrenceOE, gharibyan2020characterization, dowling_equilibration_2023, Lezama_temporal_2023, hahn2025predicting, fritzsch2025microcanonical, foini2019eigenstate, Schonle_Eigenstate_2021, Dowling2023relaxationof}. The dynamics of two-point correlations have been studied in Ref. \cite{alhambra_time_2020}, where bounds on timescales were obtained from a weak ETH assumption. Multi-time correlations were shown in Ref. \cite{dowling_equilibration_2023} to equilibrate under coarse-graining assumptions, and Markovian behavior in closed systems has been linked to the emergence of classicality via consistent histories of quantum trajectories \cite{Nation_snapshots_2020, strasberg_classicality_2023, Strasberg_classicality2_2023, Strasberg_2024}. In Ref. \cite{odonovan_quantum_2024}, the ETH was used to derive a Markovian master equation of Lindblad form. These works each suggest fundamental links between the ETH paradigm and concepts from open quantum systems, where Markovianity may emerge as an effective description of local observables in closed systems under suitable conditions. 

More complex multi-point correlations are also widely used as a tool for the study of chaos in quantum systems. The out-of-time-ordered correlator (OTOC) being a central tool, which in chaotic systems has dynamics which exponentially decays with a rate given by a quantum extension of a Lyaponov exponent \cite{xu_locality_2019}, and can be exploited as a measure of quantum chaos \cite{huang_finite-size_2019,mataprl, bhattacharyya_towards_2022, garcia2022, riddell_scaling_2023}. Notably, whilst the OTOC is a complex correlation function, experimental methods have been developed for its measurement in multiple experimental and quantum-computational settings \cite{Garttner2017, Mumford_2020, BlocherMeasuring2022, GreenExperimental2022, KastnerAncillafreemeasurement2024, Abanin2025}. Due to it's applicability as a measure of chaos, the links between the ETH/RMT and OTOCs are of great interest \cite{Roberts2017, Bergamasco2019, garcia2022, Shukla2022}. The behavior of the OTOC under chaotic conditions has motivated an extended form of the ETH where observable correlations at high orders factorize \cite{foini2019eigenstate}. However to our knowledge no approach to obtain complete dynamical behavior has yet been achieved. 

In this work we exploit an approach of chaotic wave-functions, which amount to a coarse-graining in energy of the eigenstate distributions, to calculate the full time dynamics of multi-time observable correlation functions. We obtain analytical expressions for the decay of one-, two- and four-point correlators, with the OTOC being a realisation of the latter. We show that our expressions match numerical exact diagonalisation calculations of a non-integrable quantum spin-chain. Expressions for dynamical evolution of such correlation functions open the door to a more general understanding of the emergence of Markovianity, comparing decay and system timescales directly, as well as chaotic behavior by the analytical  description of a Lyapunov exponent in certain parameter regimes.

This article is arranged as follows. First is Sec. \ref{P}, we introduce the core framework of quantum chaotic wave-functions that we will use throughout the text. Based on this in Sec. \ref{CF} we derive analytical results for one-, two-, and four-point correlation functions. In Sec. \ref{ED} we compare the prediction from RMT with the exact diagonalization of a spin chain. In Sec. \ref{CRF} we discuss some implications of our results, relating them to quantum regression of correlation functions \cite{Swain_1981, Blocher2019} and bounds on chaotic dynamics \cite{Maldacena2016}. Finally, the conclusions are presented in Sec. \ref{C}.

\section{Preliminaries}\label{P}

\subsection{Chaotic wave-functions}

One analytically tractable approach for describing quantum chaotic systems is to consider a random matrix Hamiltonian of the form $\hat{H}=\hat{H}_0 + \hat{H}_I$. The energies of the deterministic part $\hat{H}_0$ are equally spaced by $\omega$. The interaction term $\hat{H}_I$ is a random matrix sampled from Gaussian orthogonal ensemble, such that the full Hamiltonian follows the distribution $P(h)\propto \exp(-\frac{N}{4g^{2}} \mathrm{Tr}h^2)$. The matrix elements of $\hat{H}$ have zero mean, and their second moment is determined by the size of the matrix $N$ and the coupling strength $g$. This model was introduced in \cite{deutsch_quantum_1991} and is referred to as the Deutsch model. 

For convenience, the Hamiltonian is written in the eigenbasis $\{|\phi_\alpha\rangle\}_{\alpha=1}^N$ of $\hat{H}_0$, which is such that $\hat{H}_0|\phi_\alpha\rangle=E_\alpha|\phi_\alpha\rangle$, $\alpha=1,\ldots,N$, whereas for the eigenbasis of $\hat{H}$ we have $\hat{H}|\psi_\mu\rangle=E_\mu|\psi_\mu\rangle$, $\mu=1,\ldots,N$. In principle, we are able to write $|\psi_\mu\rangle=\sum_\alpha c_\mu(\alpha)|\phi_\alpha\rangle$, where $c_\mu(\alpha)$ are random variables, that we shall call random wave functions. Their properties depend on the properties of $\hat{H}_I$. Together with the orthogonality condition $\sum_{\mu\neq\nu} \langle \psi_\mu|\psi_\nu\rangle=0$, this fact forms the basis of the analysis in \cite{Nation2018}, where the approximate distribution of such coefficients is studied in the form 
\begin{equation} \label{pdf_c_main}
p(c)=\frac{1}{Z_p}\exp\left[-\sum_{\mu\alpha}\frac{c_\mu^2(\alpha)}{2\Lambda(\mu,\alpha)}\right]\prod_{\substack{\mu\nu\\\mu>\nu}}\delta\left(\sum_\alpha c_\mu(\alpha)c_\nu(\alpha)\right),
\end{equation}
with $Z_p$ being the partition function and $\Lambda(\mu,\alpha)$ a normalised smooth function peaked at $E_\mu=E_\alpha$. 
In Eq. \eqref{pdf_c_main} the orthogonality constraint accounts for correlations of the random wave functions $c_\mu(\alpha)$. Due to these correlations, non-Gaussian corrections appear when one considers ensemble averaged quantities, that are written in terms of random wave functions. For more details on the above results we refer to \cite{Nation2018, Nation2019}.

\begin{figure} 
\includegraphics[width=0.48\textwidth]{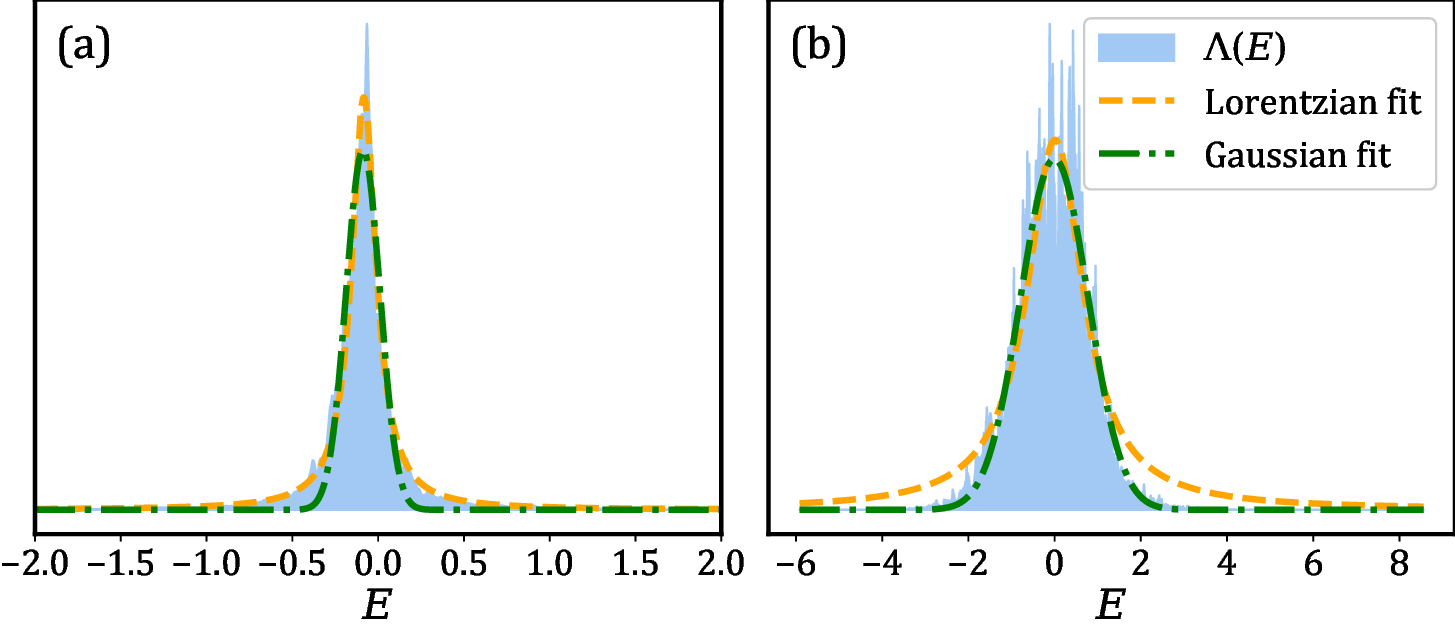}
\caption{Coarse-grained mid-energy chaotic eigenstates $\Lambda(\mu, \alpha)$ (see Eq. \eqref{eq:Lambda_def}) of the spin-chain model used in Sec. \ref{ED}, with Lorentizian and Gaussian fits in weak coupling (a) and strong coupling (b) limits. Parameters: $J_x^{\rm i} =$ 0.1 (a), 0.8 (b), each have $B_z^{\rm s}=B_x^{\rm s}=0.4$, $B_x^{\rm b} = 0.3$,  $J_x^{\rm b}=0.7$, $J_z^{\rm i}=0.2$, $r_1=5$, $r_2=10$, $N=12$.}
\label{fig1}
\end{figure}

Consistently with the framework so far, in full generality we can define chaotic eigenstate as an eigenstate fulfilling the above expansion with coefficients following the distribution \eqref{pdf_c_main}, where $\Lambda(\mu,\alpha)=\Lambda(E_\mu-E_\alpha)$ is a smooth function of $E_\mu-E_\alpha$ with maximum at $E_\mu=E_\alpha$. Moreover, $\sum_\alpha \Lambda(\mu,\alpha)=\sum_\mu \Lambda(\mu,\alpha)=1$ and
\begin{equation}\label{eq:Lambda_def}
    \langle c_\mu(\alpha)c_\nu(\beta)\rangle_V=\delta_{\mu\nu}\delta_{\alpha\beta}\Lambda(\mu,\alpha)
\end{equation}
is the two-point eigenstate correlation function. With $\langle\cdot\rangle_V$ is denoted the average, taken over the realizations of the random wave function, or equivalently over the realizations of a random Hamiltonian. This can also be understood as a coarse-graining of a single eigenstate, as seen in Fig. \ref{fig1}. For the Deutsch model, it can be shown that
\begin{equation} \label{lorentzian}
    \Lambda(\mu,\alpha) = \frac{\omega\Gamma/\pi}{(E_\mu-E_\alpha)^2+\Gamma^2},
\end{equation} 
with an energy linewidth given by $\Gamma=\pi g^2/\omega N$, which can be seen to fit well to numerical results in chaotic systems in the weak coupling regime \cite{Nation2018, Nation2019}, while in strong coupling limits numerical results often show that $\Lambda(\mu,\alpha)$ has Gaussian form \cite{santos2012onset, atas2017quantum}. We show this for the model used below in Fig. \ref{fig1} (a) and (b). Crucially, our approach in the following does not rely on the underlying RMT model, and assumes only the existence of some smooth function $\Lambda$.


In the chaotic eigenstates setting, we aim at an analytical description of the dynamics of a set of local observables $\{\hat{A}_j\}_{j=1}^M$, i.e. observables defined on a local subspace of the total Hilbert space. There are two central properties that we shall make use of: sparsity and smoothness. By sparsity, for any pair of indices $\alpha, \beta$ and for $1\leq j \leq M$ we can write 
\begin{equation}
    \langle \phi_\alpha |\hat{A}_j|\phi_\beta\rangle = \sum_{n\in N_j} \langle \phi_\alpha |\hat{A}_j|\phi_{\alpha+n}\rangle \delta_{\alpha+n,\beta},
\end{equation}
where $N_j\subset\mathbb{N}$ and $|N_j|\ll N$. This property can be seen to follow for any local observable (or a product of local observables). The smoothness property is defined as the quantity $\sum_\alpha \Lambda(\mu,\alpha)  \langle \phi_\alpha |\hat{A}_j|\phi_{\alpha+n}\rangle$ varying smoothly as a function of the energy $E_\mu$. More rigorous conditions and discussion of this property are outlined in Ref. \cite{Nation_thesis}. In fact, the latter is a necessary condition for the definition of microcanonical average of an observable that varies smoothly in energy. We thus suggest that this condition is a minimal condition necessary for quantum statistical physics to emerge in a closed system. 

In this work we use the above assumption of chaotic wave-functions to describe the dynamics of multi-point correlations in non-integrable quantum systems. As the above methodology is exploited to derive both the diagonal \cite{deutsch_quantum_1991} and off-diagonal \cite{Nation2018} ETH, the smoothness conjecture on chaotic wave-functions in non-integrable systems and it's consequences are of key importance. 

\subsection{Signatures of Markovianity and chaos in multi-time observable correlations}

Multi-time correlation functions are crucial objects in many theoretical concepts in quantum theory, with many vital properties of quantum dynamics resulting from, or understandable via, their behaviour. For example, Markovian dynamics emerges when the timescale of decay of two-point environmental correlations is much smaller than the characteristic relaxation timescale of the system \cite{Breuer_Petruccione_book}.
More generally, the second order environmental correlations can be used to obtain non-Markovian dynamics either perturbatively \cite{Breuer_Petruccione_book}, or with non-perturbative approaches such as hierarchical equations of motion \cite{Ishizaki2009}, or tensor network methods \cite{Strathearn2018}.

The properties of such two-point correlations are thus crucial for relating dynamics in closed non-integrable systems to approximations used in open quantum systems. Indeed, ETH environments have been shown to yield an effective Markovian master equation \cite{odonovan_quantum_2024}. Here we will see that a similar picture may be justified from the chaotic wave-functions approach when we consider the following results as describing the behavior of a chaotic environment.

More complicated correlation functions of system observables can be used to probe the chaoticity of quantum dynamics \cite{garcia2022}. The OTOC, defined as
\begin{equation}
    F(t) = \langle \hat{A}_1 (t) \hat{A}_2 (0)\hat{A}_1(t) \hat{A}_2 (0) \rangle,
\end{equation}
quantifies the scrambling of initial state information, via probing how the support of the Heisenberg operator $\hat{A}_1(t)$ grows in time. The relation to chaotic dynamics can be made through the related squared commutator,
\begin{align}
    \tilde{C}(t) = \langle |[\hat{A}_1(t), \hat{A}_2(0)]|^2\rangle =  C(t) + I(t) - 2\textrm{Re}[F(t)]),
\end{align}
where $D(t)=\langle \hat{A}_{2}(0)\hat{A}^{2}_{1}(t)\hat{A}_{2}(0)\rangle$ and $I(t)=\langle \hat{A}_{1}(t)\hat{A}^{2}_{2}(0)\hat{A}_{1}(t)\rangle$. This quantity describes how the commutation of two operators changes in time. For chaotic systems, this is expected to evolve in time with a `butterfly velocity', yielding a quantum generalisation of a Lyapunov exponent $\lambda_L$ describing the growth of the commutator at early times, namely $\tilde{C}(t) \sim e^{\lambda_L t}$ \cite{Maldacena2016, garcia2022}.

\section{Observable correlation functions}\label{CF}
In this section we derive the dynamics of observable correlation functions in a system with chaotic eigenstates. Consider the set of local observables $\{\hat{A}_j\}_{j=1}^M$. For $1\leq j\leq M$ the dynamics in Heisenberg picture $\hat{A}_{j}(t)=e^{i\hat{H}t}\hat{A}_{j}e^{-i\hat{H}t}$ is  
\begin{equation} \label{Heisenberg}
\hat{A}_j(t)=\sum_{\mu\nu}\sum_{\alpha\beta}c_\mu(\alpha)c_\nu(\beta)a^j_{\alpha\beta}e^{i(E_\mu-E_\nu)t}|\psi_\mu\rangle\langle\psi_\nu|,
\end{equation}
where $a_{\alpha\beta}^j = \langle\phi_\alpha|\hat{A}_j|\phi_\beta\rangle$ are the matrix elements in the non-interacting basis. 

Let $1\leq j_1,j_2,\ldots,j_k\leq M$ to be $k$ indices, where repetition is allowed. In general, an observable correlation function function defined on the set $\{\hat{A}_j\}_{j=1}^M$ is given by
\begin{equation}
\begin{split}
    \langle \hat{A}_{j_1}(t_1)\hat{A}_{j_2}(t_2)&\ldots\hat{A}_{j_k}(t_k)\rangle\\
    &=\mathrm{Tr}(\hat{\rho}\hat{A}_{j_1}(t_1)\hat{A}_{j_2}(t_2)\ldots\hat{A}_{j_k}(t_k)),
\end{split}    
\end{equation}
where $\hat{\rho}=\sum_{\alpha_{0}\beta_{0}}\rho_{\alpha_{0}\beta_{0}}|\phi_{\alpha_{0}}\rangle\langle\phi_{\beta_{0}}|$ is the initial density matrix. By using Eq. \eqref{Heisenberg}, we can write
\begin{widetext} 
    \begin{equation} \label{ocf}
    \begin{split}
        \langle \hat{A}_{j_1}(t_1)\hat{A}_{j_2}(t_2)\ldots\hat{A}_{j_k}(t_k)\rangle =&\sum_{\substack{\mu_1 \mu_2 \mu_3\ldots\\\mu_k\nu_k}} \sum_{\substack{\alpha_0\alpha_1\ldots\alpha_k\\\beta_0\beta_1\ldots\beta_k}} c_{\nu_k}(\alpha_0)c_{\mu_1}(\beta_0)c_{\mu_1}(\alpha_1)c_{\mu_2}(\beta_1)\ldots c_{\mu_k}(\alpha_k)c_{\nu_k}(\beta_k)\\
         &\times \rho_{\alpha_0\beta_0}a^{j_1}_{\alpha_1\beta_1}a^{j_2}_{\alpha_2\beta_2}\ldots a^{j_k}_{\alpha_k\beta_k}e^{i(E_{\mu_1}-E_{\mu_2})t_1}e^{i(E_{\mu_2}-E_{\mu_3})t_2}\ldots e^{i(E_{\mu_k}-E_{\nu_k})t_k}.
    \end{split} 
    \end{equation}
\end{widetext}

\subsection{Self-averaging}
The self-averaging property of large random matrices allows us to replace summations over the product of random wave functions by summations over the corresponding ensemble average \cite{Debalow_relaxation_2020, Nation2019ergodicityprobes},
\begin{equation}
    \begin{split}
        &\sum_{\substack{\mu_1 \ldots\mu_k\nu_k\\\alpha_0\ldots\alpha_k\\\beta_0\ldots\beta_k}} c_{\nu_k}(\alpha_0)c_{\mu_1}(\beta_0)\ldots c_{\mu_k}(\alpha_k)c_{\nu_k}(\beta_k)\\
        &\to\sum_{\substack{\mu_1 \ldots\mu_k\nu_k\\\alpha_0\ldots\alpha_k\\\beta_0\ldots\beta_k}} \langle c_{\nu_k}(\alpha_0)c_{\mu_1}(\beta_0)\ldots c_{\mu_k}(\alpha_k)c_{\nu_k}(\beta_k)\rangle_V.
    \end{split}
\end{equation}
Equivalently, for a system with chaotic eigenstates it holds that
\begin{equation}
\begin{split}
    \langle \hat{A}_{j_1}(t_1)\hat{A}_{j_2}(t_2)&\ldots\hat{A}_{j_k}(t_k)\rangle\\ &=\langle \hat{A}_{j_1}(t_1)\hat{A}_{j_2}(t_2)\ldots\hat{A}_{j_k}(t_k)\rangle_V.
\end{split}    
\end{equation}
The latter indicates that in such setting the observable dynamics is ensemble specific, and depends on the parameters of the random wave function distribution. Moreover, we see that seeking an expression for the observable correlation function (\ref{ocf}), we need to know the multi-point eigenstate correlation functions. 

A standard approach to the correlation function of jointly distributed random variables relies on the moment generating function (MGF). In our case, given that $N$ is the size of the matrices which represent $\hat{H}_0$ and $\hat{H}$, the MGF is 
\begin{equation} \label{definition_mgf}
    G_{\mu_1,\ldots,\mu_N}(\vec{\xi}_{\mu_1},\ldots,\vec{\xi}_{\mu_N})=\mathbb{E}\exp\Big[\sum_{j=1}^N \vec{\xi}_{\mu_j}\cdot \vec{c}_{\mu_j}\Big],
\end{equation}
and depends on at most $N$ parameters of the type $\vec{\xi}_{\mu_j}=(\xi_{\mu_j,1},\ldots,\xi_{\mu_j,N})$. The expectation in \eqref{definition_mgf} is defined with respect to the distribution \eqref{pdf_c_main}. An expression for the MGF $G_{\mu_1,\mu_2}$ has been previously derived in \cite{Nation2018}. Here we extend the result to the case of $G_{\mu_1,\ldots,\mu_N}$. We find (see Appendix \ref{appendix:mgf} for more details) 
\begin{widetext} 
    \begin{equation} 
    \begin{split}
        G_{\mu_{1}\ldots\mu_{n}}\propto \exp\Bigg{[}\sum_\alpha\sum_{i=1}^{n}\frac{\Lambda(\mu_{i},\alpha)}{2}\xi_{\mu_{i}\alpha}^2-\frac{1}{2}\sum_{\alpha\beta}\sum_{i\neq j}^{n}\xi_{\mu_{i}\alpha}\xi_{\mu_{i}\beta}\xi_{\mu_{j}\alpha}\xi_{\mu_{j}\beta}\frac{\Lambda(\mu_{i},\alpha)\Lambda(\mu_{i},\beta)\Lambda(\mu_{j},\alpha)\Lambda(\mu_{j},\beta)}{\Lambda^{(2)}(\mu_{i},\mu_{j})}\Big].
    \end{split} 
    \end{equation}
\end{widetext}
Arbitrary eigenstate correlation function can be calculated by differentiating a suitable MGF, and subsequently evaluating the derivative at zero
\begin{equation}
    \begin{split}
        \langle c_{\mu_{1}}(\alpha_{1})&\ldots c_{\mu_{n}}(\alpha_{n})\rangle_{V}\\
        &\propto\partial_{\xi_{\mu_{1},\alpha_{1}}}
\ldots\partial_{\xi_{\mu_{n},\alpha_{n}}}G_{\mu_{1}\ldots\mu_{n}}\Big|_{\vec{\xi}_{\mu_{1}}=0\ldots\vec{\xi}_{\mu_{n}}=0}.
    \end{split}
\end{equation}

Two types of terms emerge - terms corresponding to the Gaussian-like behavior of the random wave functions, and non-Gaussian corrections due to the orthogonality condition. For the multi-point correlation functions, that are used to obtain the results below, and their detailed derivation, we refer to Appendix \ref{appendix:eigenst_c_f}. 

\subsection{Analytical results}
We proceed by considering some particular observable correlation functions. Under the minimal assumptions of smoothness and sparsity, firstly we focus on the evolution of the expectation value $\langle\hat{A}_1(t)\rangle$. We define $\Omega(t):=\int \omega^{-1}\Lambda(E) e^{-iEt}\,dE$. By using four-point eigenstate correlation functions, it can be shown that (see Appendix \ref{appendix:general_ocf}) the leading-order behavior is given by (see also Refs. \cite{Nation2019, Debalow_relaxation_2020})
\begin{equation} \label{1-point-general-H}
    \langle\hat{A}_1(t)\rangle = \left(\langle \hat{A}_1(t)\rangle_{\hat{H}_0}-(A_1)_{\textrm{DE}}\right)\Omega^2(t)+(A_1)_{\textrm{DE}},
\end{equation}
where $(A_1)_{\textrm{DE}}$ is the diagonal ensemble average of the observable $\hat{A}_1$, defined by $(A_1)_{\textrm{DE}}=\textrm{Tr}(\hat{A}_1\hat{\rho}_{\textrm{DE}})$ with $\hat{\rho}_{\textrm{DE}}=\sum_\mu |b_\mu|^2 |\psi_\mu\rangle\langle\psi_\mu|$ and $b_\mu = \langle\psi_\mu|\Psi_0\rangle$, where $|\Psi_0\rangle$ is the initial state. The time evolution in the non-interacting Hamiltonian $\langle\hat{A}_1(t)\rangle_{\hat{H}_0}=\sum_{\alpha\beta}\rho_{\beta\alpha}a^1_{\alpha\beta}e^{-i(E_\alpha-E_\beta)t}$ is easily obtained for most of the systems.

Similarly, for the two-point observable correlation function we have,
\begin{equation} \label{2-point-general-H}
    \begin{split}
        \langle \hat{A}_1(t)\hat{A}_2(0)\rangle
        =&\Big(\langle \hat{A}_1(t)\hat{A}_2(0)\rangle_{\hat{H}_0}-(A_1)_{\textrm{DE}}\langle \hat{A}_2(0)\rangle\Big)\Omega^2(t)\\&+(A_1)_{\textrm{DE}}\langle \hat{A}_2(0)\rangle,
    \end{split}
\end{equation}
where $\langle\hat{A_1}(t)\hat{A}_2(0)\rangle_{\hat{H}_0}$ is the dynamics defined through the non-interacting Hamiltonian,
\begin{equation} \label{2-point-general-H0}
    \langle\hat{A_1}(t)\hat{A}_2(0)\rangle_{\hat{H}_0} = \sum_{\alpha_0\beta_0\alpha}\rho_{\alpha_0\beta_0}a^1_{\beta_0\alpha}a^2_{\alpha\alpha_0}e^{i(E_{\beta_0}-E_{\alpha})t}.
\end{equation}

This result is based on six-point eigenstate correlation functions. We find that many of the non-Gaussian corrections do not contribute significantly. The resultant dynamics is mainly determined by the correlation function involving three distinct eigenstates of $\hat{H}$.

We remark that by Eq. (\ref{2-point-general-H}) and a shift of the initial state, one can calculate two-time correlation functions of the form $\langle \hat{A_1}(t_{1})\hat{A}_2(t_{2})\rangle$, see Appendix \ref{appendix:2t}. 

The most complex observable correlation function that we are interested in, is the four-point one $\langle \hat{A_1}(t)\hat{A}_2(0)\hat{A}_3(t)\hat{A}_4(0)\rangle$, where we consider observables with zero diagonal ensemble average. In this case, the calculation simplifies considerably and boils down to the observation that the correlation function maximally spread over the eigenstates dominates the dynamics. We obtain (see Appendix \ref{appendix:otoc})
\begin{equation} \label{4-point-general-H}
    \begin{split}
        \langle \hat{A_1}(t)\hat{A}_2(0)&\hat{A}_3(t)\hat{A}_4(0)\rangle\\
        &=\langle \hat{A_1}(t)\hat{A}_2(0)\hat{A}_3(t)\hat{A}_4(0)\rangle_{\hat{H}_0}\Omega^4(t).
    \end{split}
\end{equation}

A particular case of such observable correlation function is the out-of-time-ordered correlator (OTOC) $\langle\hat{A}_1(t)\hat{A}_2(0)\hat{A}_1(t)\hat{A}_2(0)\rangle$, the contribution of highest complexity to the squared commutator $\tilde{C}(t)=\langle|[\hat{A}_1(t),\hat{A}_2(0)]|^2\rangle$ described above. The squared commutator is invariant under a shift of both observables, such that the corresponding diagonal ensemble averages are zero. This is the basis of the analysis in Appendix \ref{appendix:otoc}, where we show that in full generality $\tilde{C}(t)$ is polynomial of degree four in $\Omega(t)$, given by
\begin{widetext}
\begin{align} \label{main_text:otoc}
    \tilde{C}(t) = &\bigg(-2\langle \hat{A}^0_1(t)\hat{A}^0_2(0)\hat{A}^0_1(t)\hat{A}^0_2(0)\rangle_{\hat{H}_0}+\langle\hat{A}_1^0(t)\big((\hat{A}_2^0(0))^2\big)\hat{A}_1^0(t)\rangle_{\hat{H}_0}\bigg)\Omega^4(t)\\\notag
    +&\bigg(\langle\hat{A}^0_2(0)(\hat{A}_1^0(t))^2\hat{A}_2^0(0)\rangle_{\hat{H}_0}+\langle(\hat{A}_1^0)^2(t)\rangle_{\hat{H}_0}\big((A_2^0)^2\big)_{\textrm{DE}} -((A^0_1)^2)_{\textrm{DE}}\langle(\hat{A}_2^0(0))^2\rangle-\big((A_1^0)^2\big)_{\textrm{DE}}\big((A_2^0)^2\big)_{\textrm{DE}}\bigg)\Omega^2(t)\\\notag+&((A^0_1)^2)_{\textrm{DE}}\langle(\hat{A}_2^0(0))^2\rangle+\big((A_1^0)^2\big)_{\textrm{DE}}\big((A_2^0)^2\big)_{\textrm{DE}},
\end{align}    
\end{widetext}
where the shift applied to the observable $\hat{A}_i$ to ensure zero average is $\hat{A}_i^0=\hat{A}_i-(A_i)_{\textrm{DE}}$.

Further, given an expression for $\Lambda(E)$, one can calculate $\Omega(t)$ and characterize the time evolution of the observable correlation functions and the squared commutator. In the weak coupling regime, according to Eq. \eqref{lorentzian}, $\Lambda(E)$ has Lorentzian form, 
$
\Lambda(E)=\frac{\omega\Gamma/\pi}{E^2+\Gamma^2}.
$
Then we obtain $\Omega(t)=e^{-\Gamma t}$. The observable correlation functions approach the corresponding long time average value exponentially with rate $\Gamma$. Equations \eqref{1-point-general-H}, \eqref{2-point-general-H} and \eqref{4-point-general-H} become
\begin{widetext}
\begin{align} 
    \langle\hat{A}_1(t)\rangle &= \left(\langle \hat{A}_1(t)\rangle_{\hat{H}_0}-(A_1)_{\textrm{DE}}\right)e^{-2\Gamma t}+(A_1)_{\textrm{DE}}, \label{1-point-weak}\\
    \langle \hat{A}_1(t)\hat{A}_2(0)\rangle&=\Big(\langle \hat{A}_1(t)\hat{A}_2(0)\rangle_{\hat{H}_0}-(A_1)_{\textrm{DE}}\langle \hat{A}_2(0)\rangle\Big)e^{-2\Gamma t}+(A_1)_{\textrm{DE}}\langle \hat{A}_2(0)\rangle,\label{2-point-weak}\\
    \langle \hat{A_1}(t)\hat{A}_2(0)\hat{A}_3(t)\hat{A}_4(0)\rangle&=\langle \hat{A_1}(t)\hat{A}_2(0)\hat{A}_3(t)\hat{A}_4(0)\rangle_{\hat{H}_0}e^{-4\Gamma t}\label{4-point-weak}.
\end{align}
\end{widetext}

We note here that the result for $\langle\hat{A}_1(t)\rangle$, given by Eq. \eqref{1-point-weak}, has been previously obtained in \cite{Nation2019}, and generalised to structured random matrices in \cite{Debalow_relaxation_2020}.

As noted above, outside the perturbative regime, it is observed that the chaotic wave-functions take a Gaussian form \cite{Nation2018, atas2017quantum}, we thus write
\begin{equation}
    \Lambda(E)=\frac{\omega}{2\sqrt{\pi K}}e^{-\frac{E^2}{4K}},
\end{equation}
which is to be applied to $E$, such that $E=E'-E_\alpha$, where $E_\alpha$ is an eigenvalue of $\hat{H}_0$. Now we have $\Omega(t)=e^{-Kt^2}$, thus the decay to equilibrium is Gaussian rather than exponential. The observable correlation functions are
\begin{widetext}
\begin{align} 
    \langle\hat{A}_1(t)\rangle &= \left(\langle \hat{A}_1(t)\rangle_{\hat{H}_0}-(A_1)_{\textrm{DE}}\right)e^{-2Kt^2}+(A_1)_{\textrm{DE}}, \label{1-point-strong}\\
    \langle \hat{A}_1(t)\hat{A}_2(0)\rangle&=\Big(\langle \hat{A}_1(t)\hat{A}_2(0)\rangle_{\hat{H}_0}-(A_1)_{\textrm{DE}}\langle \hat{A}_2(0)\rangle\Big)e^{-2Kt^2}+(A_1)_{\textrm{DE}}\langle \hat{A}_2(0)\rangle,\label{2-point-strong}\\
    \langle \hat{A_1}(t)\hat{A}_2(0)\hat{A}_3(t)\hat{A}_4(0)\rangle&=\langle \hat{A_1}(t)\hat{A}_2(0)\hat{A}_3(t)\hat{A}_4(0)\rangle_{\hat{H}_0}e^{-4Kt^2}\label{4-point-strong}.
\end{align}
\end{widetext}

\section{Exact diagonalization}\label{ED}
To test the validity of the multi-time correlations presented above, we consider a one-dimensional spin chain of length $N$ with Hamiltonian $\hat{H}=\hat{H}_{\rm s}+\hat{H}_{\rm b}+\hat{H}_{\rm sb}$. The system Hamiltonian $\hat{H}_{\rm s}$ is given by
\begin{equation}
    \hat{H}_{\rm s} = B_z^{\rm s} \sigma_1^z+ B_x^{\rm s} \sigma_1^x,
\end{equation}
where the single spin forming the subsystem of interest is chosen to be the one on the first site. The rest of the spin chain we call a bath, and it is described by the Hamiltonian
\begin{equation}
    \hat{H}_{\rm b} = \sum_{n=2}^N B_x^{\rm b} \sigma_n^x +\sum_{n=2}^{N-1}J_x^{\rm b}(\sigma_{n}^{+} \sigma_{n+1}^{-}+\sigma_{n}^{-}\sigma_{n+1}^{+}),
\end{equation}
based on nearest-neighbour Ising interactions along the $x$-axis. Combining $\hat{H}_{\rm s}$ and $\hat{H}_{\rm b}$ yields the non-interacting part of the Hamiltonian $\hat{H}_{0}=\hat{H}_{\rm s}+\hat{H}_{\rm b}$. The interaction between the subsystem and the bath is governed by 
\begin{equation} \label{Hint}
    \begin{split}
        \hat{H}_{\rm sb}=J_z^{\rm i} \sigma_1^z\sigma_{r_1}^{z} + J_x^{\rm i}(\sigma_{1}^{+}\sigma_{r_1}^{-}+\sigma_{1}^{-}\sigma_{r_1}^{+})\\
        +J_z^{\rm i} \sigma_1^z\sigma_{r_2}^{z} + J_x^{\rm i}(\sigma_{1}^{+}\sigma_{r_2}^{-}+\sigma_{1}^{-}\sigma_{r_2}^{+}),
    \end{split}
\end{equation}
which represents a coupling of the subsystem to two distinct bath spins at sites $r_1$ and $r_2$, where $1<r_i\leq N$. We note that this model with the system coupled in two locations to the bath Hamiltonian is chosen as we find that the Lorentzian to Gaussian behavior of the chaotic wave-functions in this case is simply observed through via increasing the system-bath coupling. This coupling form is thus chosen for simplicity of the presentation of numerical results. The Lorentzian form in particular is straightforward to observe in many weakly coupled quantum systems \cite{Nation2019}, whereas in the strong coupling regime where Gaussian chaotic wave-functions are empirically observed, a central assumption of our approach, namely that the density of states is approximately constant over the energy width of an individual eigenstate, may be violated.

We focus on observable correlation functions of the local observables $\sigma_1^x$, $\sigma_1^z$, and $\hat{P}=\mid\uparrow_{1}\rangle\langle\uparrow_{1}\mid$. The system-dependent quantities that appear in \eqref{1-point-weak}-\eqref{4-point-weak} and \eqref{1-point-strong}-\eqref{4-point-strong}, are the diagonal ensemble averages and the decay parameters $\Gamma$, $K$. Here we study numerically the function $\Lambda(\mu, \alpha)$, which is associated to the eigenstates of the system, and perform Lorentzian and Gaussian fits to obtain the decay parameters, see Fig. \ref{fig1}(a) and (b). In each case our analytical results are shown with widths $\Gamma,\, K$ of the chaotic wave-functions obtained directly from such fits, thus numerically confirming the relation between the energy linewidth of chaotic wavefunctions and the observed decay rates of multi-time correlations.

When working with spin observables we must avoid trivial cases where two-point correlation function simply follows the one-point correlation function, which is the case whenever the initial state is an eigenstate of the initially measured observable. For example, for a system initially prepared in a N\'{e}el state $|\Psi_0\rangle=|\uparrow\rangle_{\rm s} |\downarrow\uparrow\ldots\rangle_{\rm b}$, it holds that $\langle\sigma^z_1(t)\sigma^z_1(0)\rangle=\langle\sigma^z_1(t)\rangle$. To numerically verify our analytical result in an effective way, we thus introduce to the subsystem two fields, $B_z^{\rm s}$ and $B_x^{\rm s}$, along the $z$- and the $x$-axis, and choose the initial state in the center of the spectrum of $\hat{H}_{\rm s}+\hat{H}_{\rm b}$. Another strategy with regard to the initial state is to choose at random the orientation of each spin in the chain, and subsequently to average the results for the studied quantities over many realizations of the bath state. The corresponding results are provided in Appendix \ref{appendix:random_initial_state}.

\begin{figure} 
\includegraphics[width=0.48\textwidth]{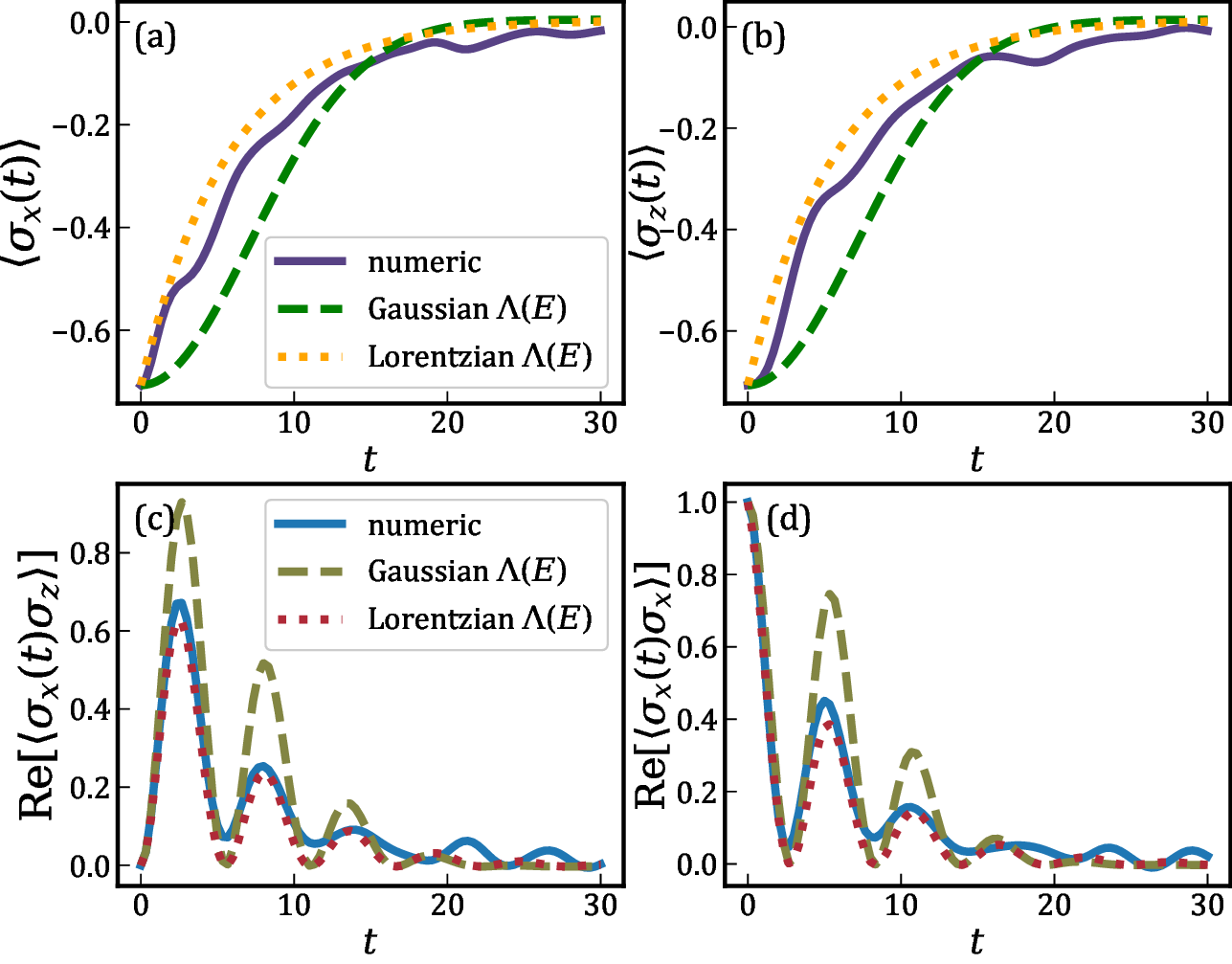}
\caption{Observable correlation functions in weak coupling regime, $J_x^{\rm i}=0.1$. (a) and (b) One-point correlation functions, analytic results given by \eqref{1-point-weak} and \eqref{1-point-strong}. (c) and (d) The real part of two-point observable correlation functions, analytic results given by \eqref{2-point-weak} and \eqref{2-point-strong}. The initial state is $|\Psi_0\rangle=|\phi_\alpha\rangle$, where $\alpha=2041$. The system consists of 12 spins and the other parameters are set to $B_z^{\rm s}=B_x^{\rm s}=0.4$, $B_x^{\rm b} = 0.3$,  $J_x^{\rm b}=0.7$, $J_z^{\rm i}=0.2$, $r_1=5$, $r_2=10$. We work with $\Gamma=0.087$ (Lorentzian $\Lambda)$ and $K=0.005$ (Gaussian $\Lambda$).}
\label{fig2}
\end{figure}

In the following analysis, we distinguish between weak and strong subsystem-bath coupling regime by controlling the value of $J_x^{\rm{i}}$. To begin, we consider weak coupling, which is characterized by Lorentzian shape of $\Lambda(\mu, \alpha)$, see Fig. \ref{fig1} (a). Then, according to the theory presented in the previous section, the integral $\Omega(t)$ leads to exponential decay to equilibrium of the observable correlation functions. Moreover, since the initial state is an eigenstate of $\hat{H}_0$, the non-interacting part of Eq. \eqref{1-point-general-H} is constant. Consequently, the behavior is anticipated to follow an exponential law entirely. This is illustrated in Fig. \ref{fig2} (a) and (b), which show the time-dependence of the expectation value for two spin observables.  

Further we focus on the general case of two observables by studying the two-point correlation function $\langle\sigma_1^x(t)\sigma_1^z(0)\rangle$, and the particular case of autocorrelation function by considering $\langle\sigma_1^x(t)\sigma_1^x(0)\rangle$, which we present in Fig. \ref{fig2} (c) and (d). Using Eq. \eqref{2-point-general-H0}, one can easily show that the non-interacting part of Eq. \eqref{2-point-general-H} has oscillatory behavior. In the weak coupling regime, as suggested by Eq. \eqref{2-point-weak}, for small times the oscillatory behavior of $\langle\sigma_1^x(t)\sigma_1^z(0)\rangle_{\hat{H}_0}$ and $\langle\sigma_1^x(t)\sigma_1^x(0)\rangle_{\hat{H}_0}$ determines the evolution, while later on the exponential decay dominates. As regards the imaginary part of two-point correlation functions, we provide analytic and numeric results in Appendix \ref{appendix:general_ocf}.

\begin{figure} 
\includegraphics[width=0.48\textwidth]{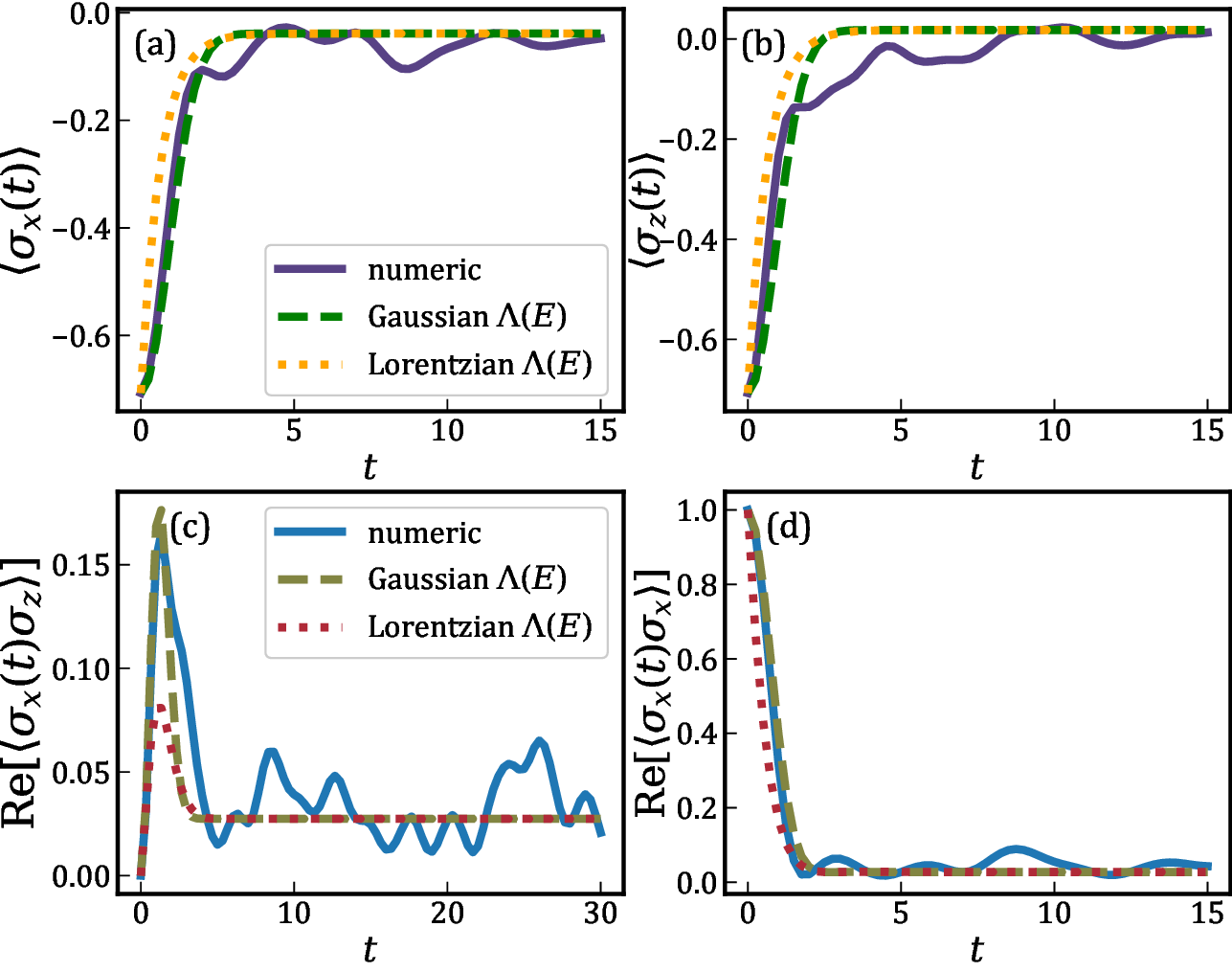}
\caption{Observable correlation functions in strong coupling regime, $J_x^{\rm i}=0.8$. (a) and (b) One-point correlation functions, analytic results given by \eqref{1-point-weak} and \eqref{1-point-strong}. (c) and (d) The real part of two-point observable correlation functions, analytic results given by \eqref{2-point-weak} and \eqref{2-point-strong}. We work with $\Gamma=0.79$ (Lorentzian $\Lambda)$ and $K=0.31$ (Gaussian $\Lambda$).}
\label{fig3}
\end{figure}

As a next test, we consider the system outside the weak coupling regime. Fig. \ref{fig1} (b) illustrates that increasing the value of $J_x^{\rm{i}}$ results in Gaussian shape of $\Lambda(\mu,\alpha)$. Consequently, the relaxation toward equilibrium is anticipated to exhibit Gaussian behavior, as dictated by $\Omega(t)$. The time dependence of the expectation values is shown in Fig. \ref{fig3} (a) and (b). We see close agreement to both the Lorentzian based Eq. \eqref{1-point-weak}, and the Gaussian based Eq. \eqref{1-point-strong}. Due to the rapid relaxation and the increased value of $\Gamma$, an effective exponential law can be identified. The Gaussian behavior is confirmed in Fig. \ref{fig3} (c), which presents the two-point observable correlation function. Although the dynamics is confined to a narrow range of values, the non-interacting part of Eq. \eqref{2-point-general-H} provides an adequate description of the long-time average. As we have seen in the perturbative regime, the non-interacting part of Eq. \eqref{2-point-general-H} contributes non-trivially to the two-point function. In contrast, in Fig. \ref{fig3} (d) we observe no oscillatory behavior, as the rapid decay toward equilibrium dominates the dynamics. Generally, in the strong coupling regime we observe stronger dynamical fluctuations at long times, which we associate to finite-size effects which are more significant in this limit.


\begin{figure} 
\includegraphics[width=0.48\textwidth]{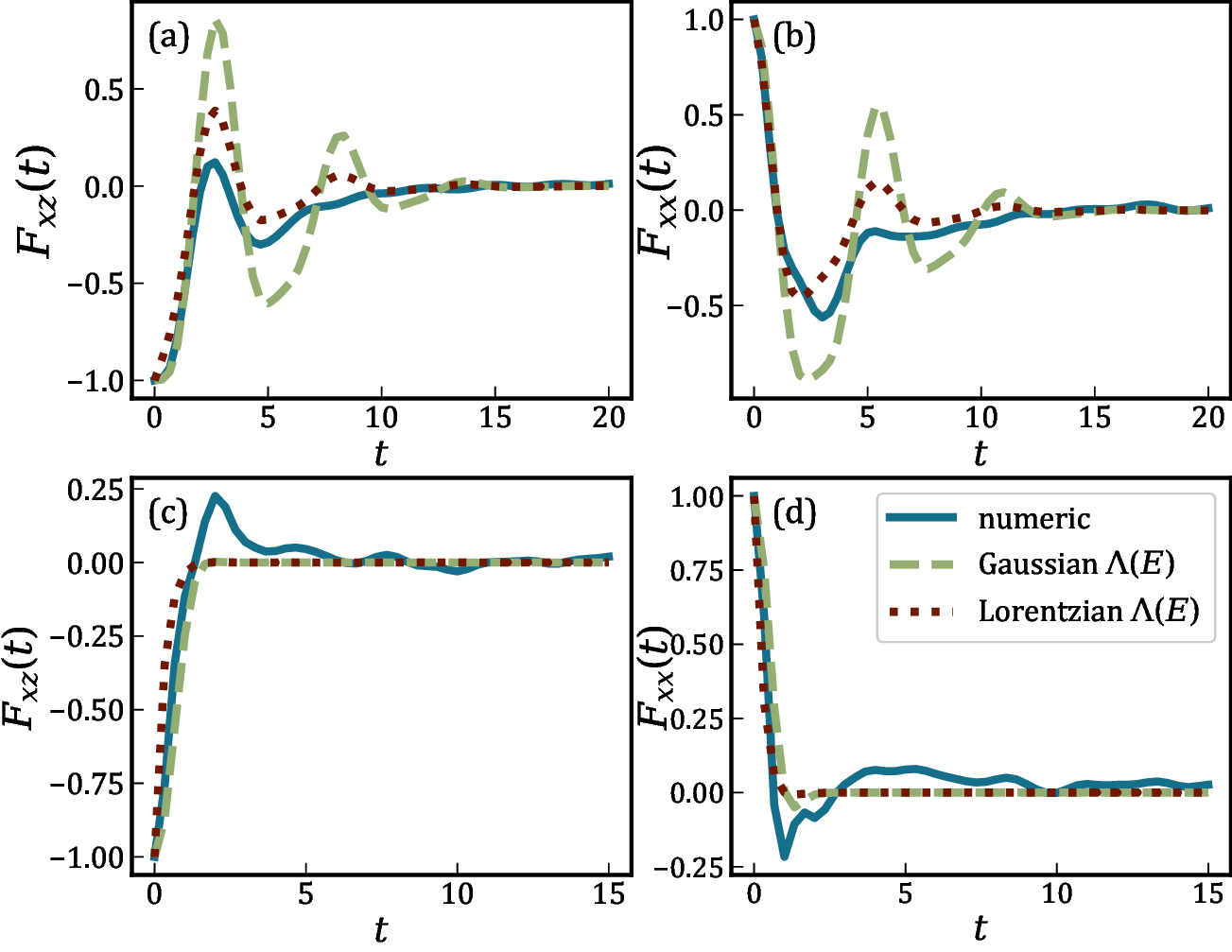}
\caption{Out-of-time-ordered correlators defined as $F_{xz}(t)=\langle\sigma_x(t)\sigma_z\sigma_x(t)\sigma_z\rangle$ and $F_{xx}(t)=\langle\sigma_x(t)\sigma_x\sigma_x(t)\sigma_x\rangle$. (a) Weak coupling regime, $J_x^{\rm i}=0.1$. (b) Strong coupling regime  $J_x^{\rm i}=0.8$. The analytic results are given by \eqref{4-point-weak} and \eqref{4-point-strong}.}
\label{fig4}
\end{figure}

Next, we investigate the applicability of Eqs. \eqref{4-point-weak} and \eqref{4-point-strong}, the four-point correlation function results. For the spin chain described above, we study two out-of-time-ordered correlators, based on the observables $\sigma_1^x$ and $\sigma_1^z$. In Fig. \ref{fig4} (a) and (b) we show the time evolution in weak coupling regime. The numerical results show complex decay behavior and compare well to the analytical prediction \eqref{4-point-weak}. Outside the weak coupling regime, Fig. \ref{fig4} (c) and (d) illustrate that both the exponential and the Gaussian law capture the decay to equilibrium of the correlation function. This is a consequence of the fast relaxation, which is well approximated by an exponential at early times even in the Gaussian case.

\begin{figure} 
\includegraphics[width=0.48\textwidth]{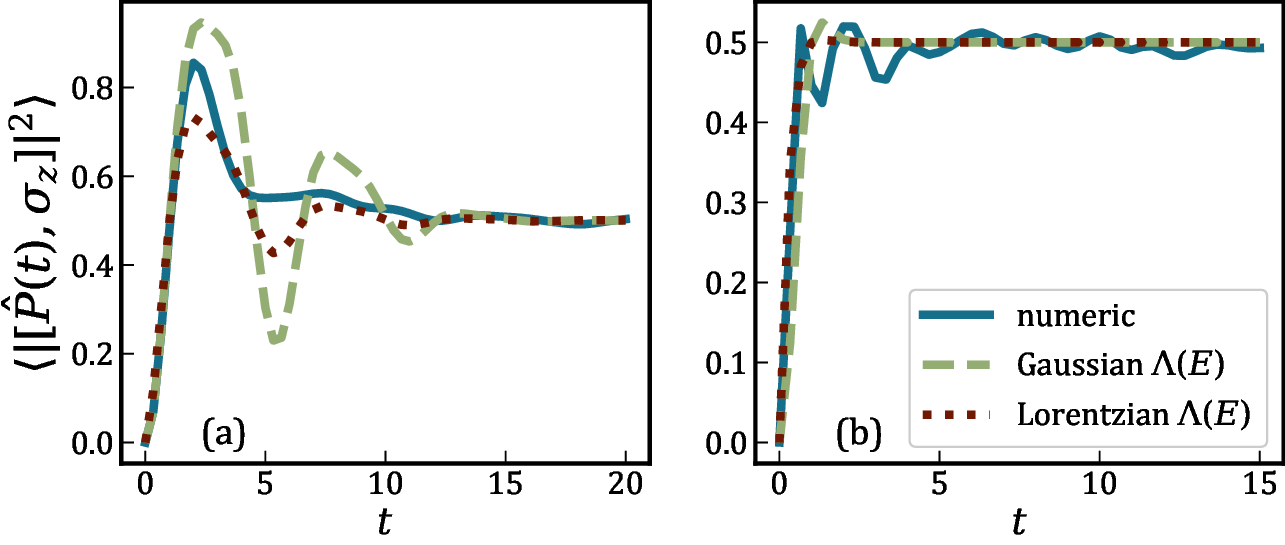}
\caption{Time dependence of the squared commutator in (a) weak ($J_x^{\rm i}=0.1$) and (b) strong ($J_x^{\rm i}=0.8$) coupling regime. The analytic results follow Eq. \eqref{commutator_ED}.}
\label{fig5}
\end{figure}

Following this, we study the squared commutator. For spin observables, the term proportional to $\Omega^2(t)$ in \eqref{main_text:otoc} vanishes. To efficiently verify the long-time average, we choose to work with the observables $\hat{P}=\mid\uparrow_{1}\rangle\langle\uparrow_{1}\mid$ and $\sigma_1^z$. Then the squared commutator takes a simple form, 
\begin{equation} \label{commutator_ED}
    \begin{split}
        \tilde{C}(t)&=\langle|[\hat{P}(t),\sigma_1^z(0)]|^2\rangle\\&=\frac{1}{2}-2\langle\hat{P}^0(t)\sigma^z_1(0)\hat{P}^0(t)\sigma^z_1(0)\rangle_{\hat{H}_0}\Omega^4(t),
    \end{split}
\end{equation}
and depends solely on the four-point correlation function. As in the preceding analysis, in the weak coupling regime the prefactor $\langle\hat{P}^0(t)\sigma^z_1(0)\hat{P}^0(t)\sigma^z_1(0)\rangle_{\hat{H}_0}$ contributes non-trivially to the time-dependence of $\tilde{C}(t)$, see Fig. \ref{fig5} (a). On the other hand, in the case of strong coupling, Fig. \ref{fig5} (b), the behavior is dominated by $\Omega(t)$, with the oscillations being suppressed and $\tilde{C}(t)\sim\Omega^4(t)$, up to constant term. As before, an effective exponential decay law is also observed. 

\section{Discussion}\label{CRF}

The results presented above demonstrate that multi-time correlation functions in chaotic quantum systems exhibit a universal decay behavior governed by the coarse-grained spectral envelope $\Lambda(E)$ of the chaotic eigenstates. Specifically, these correlations decay with powers of a single function $\Omega(t)$, the Fourier transform of $\Lambda(E)$, which depends only on system parameters and not on the choice of observable. This structure arises from the factorization of high-order eigenstate correlation functions $\langle c_\mu(\alpha) \cdots c_\nu(\beta) \rangle_V$, and closely mirrors the extended ETH conjecture proposed in Ref.~\cite{foini2019eigenstate}, where multi-point correlations of matrix elements factorize in chaotic systems.

\subsection{Quantum regression from chaos}

An conceptual consequence of our results is the emergence of a quantum regression theorem (QRT)–like structure in chaotic closed quantum systems. In open quantum systems, the QRT states that multi-time correlation functions evolve under the same dynamical generator as single-time expectation values, typically governed by a Markovian master equation. Here, we find that chaotic eigenstate statistics impose a similar structure: all local multi-time observables decay with powers of a single function $\Omega(t)$, the Fourier transform of the chaotic wave-function $\Lambda(E)$.

In the weak coupling regime, where $\Lambda(E)$ is Lorentzian,
the following regression structure holds for local observables, assuming vanishing diagonal ensemble average for simplicity $\langle \hat{A} \rangle_{\text{DE}} = 0$:
\begin{align}
    \frac{d}{dt} \langle \hat{A}(t) \rangle &= -2\Gamma \langle \hat{A}(t) \rangle, \\
    \frac{d}{dt} \langle \hat{A}_1(t) \hat{A}_2(0) \rangle &= -2\Gamma \langle \hat{A}_1(t) \hat{A}_2(0) \rangle, \\
    \frac{d}{dt} \langle \hat{A}_1(t) \hat{A}_2(0) \hat{A}_3(t) \hat{A}_4(0) \rangle &= -4\Gamma \langle \hat{A}_1(t) \hat{A}_2(0) \hat{A}_3(t) \hat{A}_4(0) \rangle.
\end{align}

Thus, one-, two-, and four-point correlation functions all decay with exponential kernels whose exponents are determined by the power of $\Omega(t)$ appearing in the analytic expressions. This directly mirrors the behavior predicted by the QRT in Markovian open quantum systems, and here emerges from the coarse-grained structure of chaotic eigenstates. This result confirms that not only do single-time observables relax thermally, but also that their multi-time correlations behave as though generated by an effective memoryless evolution, consistent with recent findings from process tensor and consistent history approaches \cite{strasberg_classicality_2023, Strasberg_classicality2_2023, Dowling2023relaxationof}.

In the strong coupling regime, where numerical results suggest a Gaussian form for $\Lambda(E)$,
the same regression structure appears, but with a time-dependent decay rate. Differentiating $\Omega^n(t)$ yields
\begin{equation}
    \frac{d}{dt} \Omega^n(t) = -2nKt \, \Omega^n(t).
\end{equation}

While the decay is no longer exponential, the structure maintains a regression-like property: higher-order correlators evolve with the same time-dependent kernel as one-point functions, up to a constant determined by the order of the correlation. This result further strengthens the interpretation of chaotic systems as effective environments, where memoryless evolution of subsystems and their observables arises from the chaotic eigenstate structure.

\subsection{OTOC timescales}

The dynamics of the out-of-time-ordered correlator (OTOC) provide a natural characterization of scrambling and information spreading in chaotic systems. In the present framework, the OTOC envelope is determined entirely by the decay kernel $\Omega(t)$, defined as the Fourier transform of the chaotic wave-function envelope $\Lambda(E)$. For example, in the simplest case of $F(0) = 1$ diagonal observables in the $\hat{H}_0$ eigenbasis,
\begin{align}
    F(t) &= \Omega^4(t), \\
    \tilde{C}(t) &= 2\big(1 - \mathrm{Re}\,F(t)\big).
\end{align}

In the weak coupling regime the full time dependence of $\tilde{C}(t)$ is dictated by the function $\Omega(t) = e^{-\Gamma t}$, which corresponds to an exponential relaxation of the OTOC envelope, or equivalently, exponential growth of the squared commutator, with rate $4\Gamma$. We to denote this scale as an effective Lyapunov rate $\lambda_L^{\mathrm{eff}} = 4\Gamma$. 
We note that for the Gaussian case, $\Omega(t) = e^{-K t^2}$, the decay is strictly Gaussian rather than exponential, though may be locally approximated by an exponential form $\tilde{C}(t) \sim e^{\lambda_{\mathrm{eff}} t}$ over a finite early-time window $t \lesssim K^{-1/2}$, with an effective rate $\lambda_{\mathrm{eff}}(t) \simeq 8Kt$. 

Finally we note that the width $\Gamma$ of the chaotic wave-function in the weak coupling case may be related to an effective temperature scale by the `eigenstate equipartition theorem' (EET) obtained in Ref \cite{Nation_snapshots_2020} for systems with quadratic system energy dispersion. This relates observable fluctuations to an effective temperature scale via an Einstein relation: $\sigma^2_X := \overline{X^2} - \overline{X}^2 \sim \beta^{-1}_{\rm eff}$, where $\overline{\cdot}$ indicates an infinite time average, and $\beta_{\rm eff}$ is an effective temperature of the initial pure state. It is also shown in Ref. \cite{Nation_snapshots_2020} (Eq. (9)) that $\sigma_X^2$ is linearly related to $\Gamma$ by a factor of order 1, and thus the scaling $\Gamma \sim \beta^{-1}_{\rm eff}$ can be inferred. We thus observe that the EET implies the scaling of the effective Lyapunov rate
    $\lambda_L^{\mathrm{eff}} = 4\Gamma \sim \beta^{-1}_{\rm eff}.$
This reproduces the temperature scaling of the Maldacena–Shenker–Stanford (MSS) \cite{Maldacena2016} bound $\lambda_L \le \frac{2\pi}{\beta}$. Note, however that the effective rate $\lambda_L^{\mathrm{\rm eff}}$ here quantifies the relaxation of local correlations, not precisely the butterfly velocity or spatial growth rate of perturbations, and is thus related but not identical to the Lyapunov exponent $\lambda_L$ of the MSS bound. We see, as with the EET, that the effective temperature assignable to \emph{individual eigenstates} of chaotic wave-functions reproduces finite temperature effects, further extending the key intuition of the ETH: thermalization occurs at the level of individual eigenstates.

\section{Conclusion}\label{C}

In this work we have exploited the theory of chaotic wave-functions \cite{Nation2018}, motivated by a random matrix theoretic approach by Deutsch to describe quantum chaotic systems \cite{deutsch_quantum_1991}, to obtain analytical descriptions of  one-, two- and four-point observable correlation functions. We find that for each correlator the dynamics is dictated by the same simple function, the Fourier transform $\Omega(t)$ of coarse-grained `chaotic wave-functions'. The decay rate is shown to depend crucially on the energy width of the chaotic wave-function. We perform numerical exact diagonalisation calculations of a quantum spin system, and explore a perturbative regime where the decay to equilibrium is exponential, and a strong-coupling regime where it is Gaussian, showing indeed the shape of the chaotic wave-function dictates the decay of observable correlation functions.

Further, a characterization of the OTOC is obtained in the form of a polynomial of degree four in $\Omega(t)$. Generally, the time-dependence of the OTOC is complex but in strong-coupling regime it might simplify as it is primarily determined by the leading order power of $\Omega(t)$. Moreover, in the specific case of spin observables $\Omega^4(t)$ determines the dynamics of the OTOC. In weak coupling regime the Lyapunov exponent is $\lambda_L=4\Gamma$, where $\Gamma$ is the decay rate consistent with all observable correlation functions. In the strong coupling regime, the behavior follows a Gaussian law, although an effective exponential law is often identifiable. For models with locally quadratic energy scaling we are able to exploit the `eigenstate equipartition theorem' \cite{Nation_snapshots_2020}, which enables the assignment of an effective temperature for individual chaotic wave-functions owing to an effective Einstein relation which is satisfied by the eigenstates themselves. This enables a relation to the Maldacena-Shenker-Stanford bound \cite{Maldacena2016} on the Lyapunov exponent, where for Lorentzian chaotic wave-functions we find the same scaling with temperature as this bound.

We have also shown the emergence of an effective quantum regression of correlation functions, in the sense that higher order correlations are each dictated by the same dynamical function $\Omega(t)$, the Fourier transform of the chaotic wave-function. Our results imply an effective Markovian description of chaotic quantum dynamics of local observables and their correlations, which yields insight into the emergence of irreversible behaviour in closed quantum systems: in the current formulation, this emerges from the effective description of chaotic eigenstates at a coarse-grained level.

We test the analytical results by means of exact diagonalization of a quantum spin chain. We study both the weak and the strong coupling regime, characterised by Lorentzian and Gaussian chaotic wave-functions, respectively, for multiple observables. We note that whilst there is in general good agreement between the analytical and numerical results, for higher order correlations a larger deviation from the analytical results is observed. We associate this to the more sensitive behaviour of higher order correlations to deviations from the assumed form of the chaotic wave-function. A problem left open is the extension of the regression theorem to generic order correlation functions.

Together, these results on two- and four-point correlations support a unifying view of chaotic quantum systems as described by the chaotic wave-function approach as both fast scramblers and effective thermal baths. The presence of a single decay function $\Omega(t)$ that governs all local multi-time correlators enables a compact analytical characterization of both thermalization and scrambling. This structure offers a potential diagnostic for chaos: deviations from these universal decay forms would indicate non-chaotic behavior, such as integrability or localization. Furthermore, the emergence of a regression-like structure implies that isolated chaotic systems can mimic the dynamics of Markovian open systems, without requiring an external environment. Our findings thus deepen the connection between quantum chaos, thermalization, and the structure of multi-time observables, and suggest new routes toward a fully microscopic theory of quantum thermodynamics in closed systems.

\section*{Acknowledgments}
Y.R.C. and P.A.I. acknowledge the Bulgarian national plan for recovery and resilience, contract BG-RRP-2.004-0008-C01 (SUMMIT: Sofia University Marking Momentum for Innovation and Technological Transfer), project number 3.1.4. C.N acknowledges funding from the EPSRC quantum
career development grant EP/W028301/1 and the
EPSRC Standard Research grant EP/Z534250/1. C.N would like to thank Diego Porras for enlightening conversations.

\bibliographystyle{apsrev4-1}
\bibliography{bibli}

\appendix

\begin{widetext}
\section{Calculation of the moment generating function} \label{appendix:mgf}

Following \cite{Nation2018}, the probability distribution of the random wave functions $c_\mu(\alpha)$ is
\begin{equation} \label{pdf_c}
p(c)=\frac{1}{Z_p}\exp\left[-\sum_{\mu\alpha}\frac{c_\mu^2(\alpha)}{2\Lambda(\mu,\alpha)}\right]\prod_{\substack{\mu\nu\\\mu>\nu}}\delta\left(\sum_\beta c_\mu(\beta)c_\nu(\beta)\right), \end{equation}
where the partition function $Z_p$ is
\[
Z_p = (2\pi)^{N^2-N/2}\Big{(}\prod_{\mu\alpha}(\Lambda(\mu,\alpha))^{1/2}\Big{)}\Bigg{(}\prod_{\substack{\mu\nu\\\mu>\nu}}\Big{(}\sum_\beta\Lambda(\mu,\beta)\Lambda(\nu,\beta)\Big{)}^{-1/2}\Bigg{)}.
\]

Given the moment generating function (MGF) $G_{\mu_1,\ldots,\mu_N}(\vec{\xi}_{\mu_1},\ldots,\vec{\xi}_{\mu_N})$ defined with respect to \eqref{pdf_c}, one can calculate eigenstate correlation functions by differentiation:
\begin{equation} \label{derivative}
    \begin{split}
        \langle c_{\mu_1}(\alpha^1_{1})\ldots c_{\mu_1}(\alpha^1_{m_1})\ldots c_{\mu_n}(\alpha^n_1)\ldots &c_{\mu_n}(\alpha^n_{m_n})\rangle_V=\\
        &=\frac{1}{G_{\mu_1\ldots\mu_n}} \partial_{\xi_{\mu_1 \alpha^1_1}}\ldots\partial_{\xi_{\mu_1 \alpha^1_{m_1}}}\ldots \partial_{\xi_{\mu_n \alpha^n_1}}\ldots\partial_{\xi_{\mu_n \alpha^n_{m_n}}}G_{\mu_1\ldots\mu_n}\Bigg{|}_{\vec{\xi}_{\mu_1},\ldots,\vec{\xi}_{\mu_n}=0}.
    \end{split}
\end{equation} 

In the case of two parameters $\vec{\xi}_\mu$ and $\vec{\xi}_\nu$, as shown in \cite{Nation2018}, the MGF is 
\begin{equation} \label{mgf2}
    G_{\mu\nu}(\vec{\xi}_\mu,\vec{\xi}_\nu)\propto\exp\Bigg{[}\sum_\alpha\Big{(}\frac{\Lambda(\mu,\alpha)}{2}\xi_{\mu\alpha}^2+\frac{\Lambda(\nu,\alpha)}{2}\xi_{\nu\alpha}^2\Big{)}-\frac{1}{2}\sum_{\alpha\beta}\xi_{\mu\alpha}\xi_{\mu\beta}\xi_{\nu\alpha}\xi_{\nu\beta}\frac{\Lambda(\mu,\alpha)\Lambda(\mu,\beta)\Lambda(\nu,\alpha)\Lambda(\nu,\beta)}{\Lambda^{(2)}(\mu,\nu)}\Bigg{]}.
\end{equation}

For clarity, let us firstly consider the case of three types of random wave functions: $c_{\mu}$, $c_{\nu}$, and $c_{\rho}$. Using the distribution (\ref{pdf_c}), we write the general expression for the MGF 
\begin{equation} \label{general_mgf}
    \begin{split} 
        G_{\mu\nu\rho}(\vec{\xi_\mu},\vec{\xi_\nu},\vec{\xi_\rho})=\iiint\exp&\Bigg[-\sum_\alpha\left(\frac{c^2_\mu(\alpha)}{2\Lambda(\mu,\alpha)}+\frac{c^2_\nu(\alpha)}{2\Lambda(\nu,\alpha)}+\frac{c^2_\rho(\alpha)}{2\Lambda(\rho,\alpha)}-\xi_{\mu\alpha}c_\mu(\alpha)-\xi_{\nu\alpha}c_\nu(\alpha)-\xi_{\rho\alpha}c_\rho(\alpha)\right)\Bigg]\\
        & \times \delta(\sum_\alpha c_\mu(\alpha)c_\nu(\alpha))\delta(\sum_\alpha c_\mu(\alpha)c_\rho(\alpha))\delta(\sum_\alpha c_\nu(\alpha)c_\rho(\alpha))\prod_\alpha dc_\mu(\alpha)\,dc_\nu(\alpha)\,dc_\rho(\alpha).
    \end{split}
\end{equation} 

After expressing the Dirac delta function in Fourier form, \eqref{general_mgf} becomes
\begin{align*} 
G_{\mu\nu\rho}(\vec{\xi_\mu},\vec{\xi_\nu},\vec{\xi_\rho})&=\iiint\iiint\exp\left[-\sum_\alpha\left(\frac{c^2_\mu(\alpha)}{2\Lambda(\mu,\alpha)}+\frac{c^2_\nu(\alpha)}{2\Lambda(\nu,\alpha)}+\frac{c^2_\rho(\alpha)}{2\Lambda(\rho,\alpha)}-\xi_{\mu\alpha}c_\mu(\alpha)-\xi_{\nu\alpha}c_\nu(\alpha)-\xi_{\rho\alpha}c_\rho(\alpha)\right)\right]\\
& \times \exp\left[i\,\sum_\alpha\left(\lambda c_\mu(\alpha)c_\nu(\alpha)+\lambda' c_\mu(\alpha)c_\rho(\alpha)+\lambda''c_\nu(\alpha)c_\rho(\alpha)\right)\right]\prod_\alpha dc_\mu(\alpha)\,dc_\nu(\alpha)\,dc_\rho(\alpha)\,d\lambda\,d\lambda'\,d\lambda''.
\end{align*}

Further we shall evaluate the Gaussian integral in terms of its multivariate generalization, so that 
\begin{equation*}
    \begin{split}
        G_{\mu\nu\rho}\propto\iiint\int\exp\Bigg(-\frac{1}{2}x^T\cdot A\cdot x + J^T\cdot x\Bigg)\,dx\,d\lambda\,d\lambda'\,d\lambda''\propto [\textrm{det}(A)]^{-1/2}\iiint \exp\Bigg(\frac{1}{2} J^T\cdot A^{-1}\cdot J\Bigg)\,d\lambda\,d\lambda'\,d\lambda'',
    \end{split}
\end{equation*}
where we have introduced the $3N$-dimensional vectors $x=[c_\mu(1), c_\nu(1), c_\rho(1),\ldots,c_\mu(N), c_\nu(N), c_\rho(N)]^\textrm{T}$ and $J=[\xi_{\mu,1}, \xi_{\nu,1}, \xi_{\rho,1},\ldots,\xi_{\mu, N},\xi_{\nu,N},\xi_{\rho,N}]^\textrm{T}$. The matrix $A$ is block-diagonal and the blocks are given by
\[
A_\alpha = 
\begin{bmatrix}
\frac{1}{\Lambda(\mu,\alpha)}& -i\lambda& -i\lambda'\\
-i\lambda&\frac{1}{\Lambda(\nu,\alpha)}&-i\lambda''\\
-i\lambda'&-i\lambda''&\frac{1}{\Lambda(\rho,\alpha)}
\end{bmatrix}, \hspace{15pt} 1\leq \alpha\leq N.
\]

Then for the determinant $\det{A}$ we have
\begin{align*}
(\det{A})^{1/2}=\prod_\alpha\Big[1+2i\lambda\lambda'\lambda''\Lambda(\mu,\alpha)\Lambda(\nu,\alpha)\Lambda(\rho,\alpha)+\lambda''^2\Lambda(\nu,\alpha)\Lambda(\rho,\alpha)+\lambda'^2\Lambda(\mu,\alpha)\Lambda(\rho,\alpha)+\lambda^2\Lambda(\mu,\alpha)\Lambda(\nu,\alpha)]\Big]^{1/2}
\\\times\prod_\beta\big[\Lambda(\mu,\beta)\Lambda(\nu,\beta)\Lambda(\rho,\beta)\big]^{-1/2}\\ \approx \exp\Big\{\frac{1}{2}\sum_\alpha\Big[\lambda''^2\Lambda(\nu,\alpha)\Lambda(\rho,\alpha)+\lambda'^2\Lambda(\mu,\alpha)\Lambda(\rho,\alpha)+\lambda^2\Lambda(\mu,\alpha)\Lambda(\nu,\alpha)\Big]\Big\}\prod_\beta\big[\Lambda(\mu,\beta)\Lambda(\nu,\beta)\Lambda(\rho,\beta)\big]^{-1/2}],
\end{align*}
where we have used that $\Lambda(\mu,\alpha)\Lambda(\nu,\alpha)\Lambda(\rho,\alpha)\propto (\omega/\Gamma)^3\ll 1$ in order to neglect the corresponding term, and further we have applied $\ln(1+x)\approx x$, which holds true for large $N$.

In the limit $\omega/\Gamma\ll 1$, we also obtain
\begin{align*}
    J^T\cdot A^{-1}\cdot J\approx\sum_\alpha\Big[\Lambda(\mu,\alpha)\xi^2_{\mu\alpha}+\Lambda(\nu,\alpha)\xi^2_{\nu\alpha}+\Lambda(\rho,\alpha)\xi^2_{\rho\alpha}-2i\lambda\Lambda(\mu,\alpha)\Lambda(\nu,\alpha)\xi_{\mu\alpha}\xi_{\nu\alpha}\\-2i\lambda'\Lambda(\mu,\alpha)\Lambda(\rho,\alpha)\xi_{\mu\alpha}\xi_{\rho\alpha}-2i\lambda''\Lambda(\nu,\alpha)\Lambda(\rho,\alpha)\xi_{\nu\alpha}\xi_{\rho\alpha}\Big].
\end{align*}

Given the above results, integration over $\lambda$, $\lambda'$ and $\lambda''$ yields 
\begin{equation} \label{mgf3}
    \begin{split}
        G_{\mu\nu\rho}\propto \exp\Bigg{[}\sum_\alpha\Big{(}\frac{\Lambda(\mu,\alpha)}{2}\xi_{\mu\alpha}^2+\frac{\Lambda(\nu,\alpha)}{2}\xi_{\nu\alpha}^2+\frac{\Lambda(\rho,\alpha)}{2}\xi_{\rho\alpha}^2-\frac{1}{2}\sum_{\alpha\beta}\xi_{\mu\alpha}\xi_{\mu\beta}\xi_{\nu\alpha}\xi_{\nu\beta}\frac{\Lambda(\mu,\alpha)\Lambda(\mu,\beta)\Lambda(\nu,\alpha)\Lambda(\nu,\beta)}{\Lambda^{(2)}(\mu,\nu)}\\
        -\frac{1}{2}\sum_{\alpha\beta}\xi_{\mu\alpha}\xi_{\mu\beta}\xi_{\rho\alpha}\xi_{\rho\beta}\frac{\Lambda(\mu,\alpha)\Lambda(\mu,\beta)\Lambda(\rho,\alpha)\Lambda(\rho,\beta)}{\Lambda^{(2)}(\mu,\rho)}-\frac{1}{2}\sum_{\alpha\beta}\xi_{\nu\alpha}\xi_{\nu\beta}\xi_{\rho\alpha}\xi_{\rho\beta}\frac{\Lambda(\nu,\alpha)\Lambda(\nu,\beta)\Lambda(\rho,\alpha)\Lambda(\rho,\beta)}{\Lambda^{(2)}(\nu,\rho)}\Big].
    \end{split}
\end{equation}

In full generality, one can be interested in the MGF $G_{\mu_{1},\ldots\mu_{n}}(\vec{\xi}_{\mu_{1}},\ldots\vec{\xi}_{\mu_{n}})$ for $2\leq n\leq N$. As the calculations, involving the matrix $A$ become lengthy, here we emphasize that in the case of $n=3$ and $\omega/\Gamma\ll 1$ new type of terms does not emerge.  It is not hard to see that the trend is preserved for $n>3$. The latter allows us to come to conclusions about the procedure and the form of $G_{\mu_{1},\ldots\mu_{n}}(\vec{\xi}_{\mu_{1}},\ldots\vec{\xi}_{\mu_{n}})$ for an arbitrary $n$. 

Now we have $n(n-1)/2$ Dirac delta function terms stemming from the orthogonality condition:
\begin{equation} \label{general_mgf1}
    \begin{split} 
        &G_{\mu_{1},\ldots\mu_{n}}(\vec{\xi}_{\mu_{1}},\ldots\vec{\xi}_{\mu_{n}})=\iint\ldots\int\exp\left[-\sum_{i=1}^{n}\sum_\alpha\left(\frac{c^2_{\mu_{i}}(\alpha)}{2\Lambda(\mu_{i},\alpha)}-\xi_{{\mu_{i}}\alpha}c_{\mu_{i}}(\alpha)\right)\right]\prod_{i\neq j}\delta(\sum_\alpha c_{\mu_{i}}(\alpha)c_{\mu_{j}}(\alpha))\prod_{i\alpha} dc_{\mu_{i}}(\alpha).
    \end{split}
\end{equation} 
Writing the Fourier form of the Dirac delta function leads to
\begin{equation}
\begin{split}
G_{\mu_{1},\ldots\mu_{n}}(\vec{\xi}_{\mu_{1}},\ldots\vec{\xi}_{\mu_{n}})\propto\iiint\ldots\int\exp\Bigg(-\frac{1}{2}x^T\cdot A\cdot x + J^T\cdot x\Bigg)\,dx\prod_{i=1}^{n}d\lambda_{i}\\\propto [\textrm{det}(A)]^{-1/2}\iint\ldots\int\exp\Bigg(\frac{1}{2} J^T\cdot A^{-1}\cdot J\Bigg)\prod_{i=1}^{n}d\lambda_{i}.
\end{split}
\end{equation}
Performing integration over $\lambda_{i}$ yields
\begin{equation} \label{mgf4}
    \begin{split}
        G_{\mu_{1}\ldots\mu_{n}}\propto \exp\Bigg{[}\sum_\alpha\sum_{i=1}^{n}\frac{\Lambda(\mu_{i},\alpha)}{2}\xi_{\mu_{i}\alpha}^2-\frac{1}{2}\sum_{\alpha\beta}\sum_{i\neq j}^{n}\xi_{\mu_{i}\alpha}\xi_{\mu_{i}\beta}\xi_{\mu_{j}\alpha}\xi_{\mu_{j}\beta}\frac{\Lambda(\mu_{i},\alpha)\Lambda(\mu_{i},\beta)\Lambda(\mu_{j},\alpha)\Lambda(\mu_{j},\beta)}{\Lambda^{(2)}(\mu_{i},\mu_{j})}\Big].
    \end{split}
\end{equation}

\section{Eigenstate correlation functions}
\label{appendix:eigenst_c_f}

The two-point correlation function is given by $\langle c_\mu(\alpha)c_\nu(\beta)\rangle_V=\Lambda(\mu,\alpha)\delta_{\mu\nu}\delta_{\alpha\beta}$, which corresponds to the correlation function of independent Gaussian random variables with variance $\Lambda(\mu,\alpha)$. Four-point correlation functions, previously calculated in \cite{Nation2018}, are non-null solely in the case of $\langle c_\mu(\alpha)c_\mu(\beta)c_\mu(\alpha')c_\mu(\beta')\rangle_V$ and $\langle c_\mu(\alpha)c_\mu(\beta)c_\nu(\alpha')c_\nu(\beta')\rangle_V$. The latter type can be represented as the product of two-point correlation functions with the addition of non-Gaussian correction of the form
\begin{equation} 
\langle\langle c_\mu(\alpha)c_\mu(\beta)c_\nu(\alpha')c_\nu(\beta')\rangle\rangle_V := - \frac{\Lambda(\mu,\alpha)\Lambda(\mu,\beta)\Lambda(\nu,\alpha')\Lambda(\nu,\beta')}{\sum_\gamma \Lambda(\mu,\gamma)\Lambda(\nu,\gamma)}(\delta_{\alpha\alpha'}\delta_{\beta\beta'}+\delta_{\alpha\beta'}\delta_{\beta\alpha'}),
\label{non-gaussian}
\end{equation}
so that
\begin{equation} \label{ecf4}
    \langle c_\mu(\alpha)c_\mu(\beta)c_\nu(\alpha')c_\nu(\beta')\rangle_V=\langle c_\mu(\alpha)c_\mu(\beta)\rangle_V\langle c_\nu(\alpha')c_\nu(\beta')\rangle_V + \langle\langle c_\mu(\alpha)c_\mu(\beta)c_\nu(\alpha')c_\nu(\beta')\rangle\rangle_V .
\end{equation}

In our work, we are interested in six-point correlation functions. From Eq. \eqref{derivative} once applied to the moment generating function (\ref{mgf2}) and once to \eqref{mgf3}, we obtain
\begin{align} \label{ecf6:2}
    \langle &c_\mu(\alpha_0)c_\mu(\beta_0)c_\nu(a)c_\nu(\beta)c_\nu(\alpha')c_\nu(\beta')\rangle_V=\nonumber\\
    &=\langle c_\mu(\alpha_0)c_\mu(\beta_0)\rangle_V\,\Big[\langle c_\nu(\alpha)c_\nu(\beta)\rangle_V\langle c_\nu(\alpha')c_\nu(\beta')\rangle_V +\nonumber\\
    &\hspace{4.45cm}+ \langle c_\nu(\alpha')c_\nu(\beta)\rangle_V\langle c_\nu(\alpha)c_\nu(\beta')\rangle_V 
    +\langle c_\nu(\alpha)c_\nu(\alpha')\rangle_V\langle c_\nu(\beta)c_\nu(\beta')\rangle_V\Big]\\
    &+\langle c_\nu(\alpha)c_\nu(\beta)\rangle_V\,\langle\langle c_\mu(\alpha_0)c_\mu(\beta_0)c_\nu(\alpha')c_\nu(\beta')\rangle\rangle_V +\langle c_\nu(\alpha)c_\nu(\alpha')\rangle_V\,\langle\langle c_\mu(\alpha_0)c_\mu(\beta_0)c_\nu(\beta)c_\nu(\beta')\rangle\rangle_V\nonumber\\ 
    &+\langle c_\nu(\alpha)c_\nu(\beta')\rangle_V\,\langle\langle c_\mu(\alpha_0)c_\mu(\beta_0)c_\nu(\beta)c_\nu(\alpha')\rangle\rangle_V +\langle c_\nu(\beta)c_\nu(\alpha')\rangle_V\,\langle\langle c_\mu(\alpha_0)c_\mu(\beta_0)c_\nu(\alpha)c_\nu(\beta')\rangle\rangle_V\nonumber\\
    &+\langle c_\nu(\beta)c_\nu(\beta')\rangle_V\,\langle\langle c_\mu(\alpha_0)c_\mu(\beta_0)c_\nu(\alpha)c_\nu(\alpha')\rangle\rangle_V 
    +\langle c_\nu(\alpha')c_\nu(\beta')\rangle_V\,\langle\langle c_\mu(\alpha_0)c_\mu(\beta_0)c_\nu(\alpha)c_\nu(\beta)\rangle\rangle_V\nonumber
\end{align}
and
\begin{align}\label{ecf6:3}
    \langle c_\mu(\alpha_0)c_\mu(\beta_0)c_\nu(\alpha)c_\nu(\beta)c_\rho(\alpha')c_\rho(\beta')\rangle_V=
    \langle c_\mu(\alpha_0)c_\mu(\beta_0)\rangle_V\langle c_\nu(\alpha)c_\nu(\beta)\rangle_V\langle c_\rho(\alpha')c_\rho(\beta')\rangle_V  \nonumber&\\
    +\langle c_\mu(\alpha_0)c_\mu(\beta_0)\rangle_V\,\langle\langle c_\nu(\alpha)c_\nu(\beta)c_\rho(\alpha')c_\rho(\beta')\rangle\rangle_V&\\
    +\langle c_\nu(\alpha)c_\nu(\beta)\rangle_V\,\langle\langle c_\mu(\alpha_0)c_\mu(\beta_0)c_\rho(\alpha')c_\rho(\beta')\rangle\rangle_V\nonumber&\\
    +\langle c_\rho(\alpha')c_\rho(\beta')\rangle_V\,\langle\langle c_\mu(\alpha_0)c_\mu(\beta_0)c_\nu(\alpha)c_\nu(\beta)\rangle\rangle_V,\nonumber
\end{align}
where the non-Gaussian corrections $\langle\langle \cdot \rangle\rangle_V$ are given by (\ref{non-gaussian}).

\section{Observable correlation functions} \label{appendix:general_ocf}

In this section we calculate the contributions to the observable correlation functions of interest by using the results, stated at Appendix \ref{appendix:eigenst_c_f}. 

Consider the local observable $\hat{A}_1$. For the dynamics of the expectation value we have $\langle \hat{A}_1(t)\rangle=\langle \hat{A}_1(t)\rangle_V$, and  by Eq. \eqref{ocf}
\begin{equation} \label{A1_g}
    \langle \hat{A}_1(t)\rangle=\sum_{\mu\nu}\sum_{\substack{\alpha_0\alpha\\\beta_0\beta}}\rho_{\alpha_0\beta_0}a^1_{\alpha\beta}\langle c_\nu(\alpha_0)c_\mu(\beta_0)c_\mu(\alpha)c_\nu(\beta)\rangle_Ve^{i(E_\mu-E_\nu)t}.
\end{equation}

Using Eq. \eqref{ecf4}, for the time dependent part of \eqref{A1_g}, denoted by $\langle \Delta A_1(t)\rangle$, we obtain 
\begin{equation} \label{A1_g:t}
\begin{split}
\langle \Delta A_1(t)\rangle&=\sum_{\substack{\mu\nu\\\mu\neq\nu}}\sum_{\alpha\beta}\rho_{\beta\alpha}a^1_{\alpha\beta}\Lambda(\mu,\alpha)\Lambda(\nu,\beta)e^{i(E_\mu-E_\nu)t}\\
&-\sum_{\substack{\mu\nu\\\mu\neq\nu}}\sum_{\alpha_0\alpha}\rho_{\alpha_0\alpha_0}a^1_{\alpha\alpha}\frac{\Lambda(\mu,\alpha_0)\Lambda(\mu,\alpha)\Lambda(\nu,\alpha_0)\Lambda(\nu,\alpha)}{\sum_\gamma\Lambda(\mu,\gamma)\Lambda(\nu,\gamma)}e^{i(E_\mu-E_\nu)t}\\
&-\sum_{\substack{\mu\nu\\\mu\neq\nu}}\sum_{\alpha\beta}\rho_{\alpha\beta}a^1_{\alpha\beta}\frac{\Lambda(\mu,\alpha)\Lambda(\mu,\beta)\Lambda(\nu,\alpha)\Lambda(\nu,\beta)}{\sum_\gamma\Lambda(\mu,\gamma)\Lambda(\nu,\gamma)}e^{i(E_\mu-E_\nu)t}.
\end{split}
\end{equation}

We proceed by replacing the summation of the type $\sum_\mu$ with integration via $\sum_\mu\to\int dE\,D(E)$, where $D(E)=\sum_\mu \delta(E-E_\mu)$ is the density of states. As by definition $\Lambda(\mu,\alpha)=\Lambda(E_\mu-E_\alpha)$, we introduce the notation 
\begin{equation}
   C_\alpha(t)=\int dE\,D(E)\Lambda(E-E_\alpha)e^{iEt},\hspace{25pt}C_\alpha(-t)=\int dE\,D(E)\Lambda(E-E_\alpha)e^{-iEt}.
\end{equation}
Observe that in the case when $D(E)$ is constant over the values of $E$, since $\Lambda(E-E_\alpha)$ is a symmetric function, one can write 
\begin{equation} \label{C_to_Omega}
    C_\alpha(t)=e^{iE_\alpha t}\Omega(t),\hspace{25pt}C_\alpha(-t)=e^{-iE_\alpha t}\Omega(t),
\end{equation} 
where $\Omega(t)=\int d(E-E_\alpha)\,D(E)\Lambda(E-E_\alpha)e^{i(E-E_\alpha)t}$ is a symmetric function of $t$.

Now, omitting the contribution $\mu=\nu$ to the long time average, for the first term in \eqref{A1_g:t} we have 
\begin{equation}
\sum_{\substack{\mu\nu\\\mu\neq\nu}}\sum_{\alpha\beta}\rho_{\beta\alpha}a^1_{\alpha\beta}\Lambda(\mu,\alpha)\Lambda(\nu,\beta)e^{i(E_\mu-E_\nu)t}=\sum_{\alpha\beta}\rho_{\beta\alpha}a^1_{\alpha\beta}C_\alpha(t)C_\beta(-t).
\end{equation}

Further, consider $[a^1_{\alpha\alpha}]_\mu:=\sum_\alpha a^1_{\alpha\alpha}\Lambda(\mu,\alpha)$. By the smoothness property, $[a^1_{\alpha\alpha}]_\mu$ slowly varies with $\mu$. Then we are justified in writing $\sum_\alpha a^1_{\alpha\alpha}\Lambda(\mu,\alpha)\Lambda(\nu,\alpha)\approx [a^1_{\alpha\alpha}]_{\bar{\mu}}\sum_\alpha \Lambda(\mu,\alpha)\Lambda(\nu,\alpha)$, where $[a^1_{\alpha\alpha}]_{\bar{\mu}}$ is defined with respect to the energy $E_{\bar{\mu}}=\frac{1}{2}(E_\mu+E_\nu)$. Applying this reasoning to the second term in \eqref{A1_g:t}, we get
\begin{equation}
    \begin{split}
        &\sum_{\substack{\mu\nu\\\mu\neq\nu}}\sum_{\alpha_0\alpha}\rho_{\alpha_0\alpha_0}a^1_{\alpha\alpha}\frac{\Lambda(\mu,\alpha_0)\Lambda(\mu,\alpha)\Lambda(\nu,\alpha_0)\Lambda(\nu,\alpha)}{\sum_\gamma\Lambda(\mu,\gamma)\Lambda(\nu,\gamma)}e^{i(E_\mu-E_\nu)t}\\&\approx\sum_{\substack{\mu\nu\\\mu\neq\nu}}\sum_{\alpha_0}\rho_{\alpha_0\alpha_0}[a^1_{\alpha\alpha}]_{\bar{\mu}}\frac{\Lambda(\mu,\alpha_0)\Lambda(\nu,\alpha_0)\sum_\alpha\Lambda(\mu,\alpha)\Lambda(\nu,\alpha)}{\sum_\gamma\Lambda(\mu,\gamma)\Lambda(\nu,\gamma)}e^{i(E_\mu-E_\nu)t}\\&=\sum_{\substack{\mu\nu\\\mu\neq\nu}}\sum_{\alpha_0}\rho_{\alpha_0\alpha_0}[a^1_{\alpha\alpha}]_{\bar{\mu}}\Lambda(\mu,\alpha_0)\Lambda(\nu,\alpha_0)e^{i(E_\mu-E_\nu)t}\approx\sum_{\alpha_0}\rho_{\alpha_0\alpha_0}[a^1_{\alpha\alpha}]_{\alpha_0}C_{\alpha_0}(t)C_{\alpha_0}(-t),
    \end{split}
\end{equation}
where for the last step we have used that $\sum_{\mu, \nu} [a_{\alpha\alpha}^1]_{\overline{\mu}}\Lambda(\mu, \alpha_0) \Lambda(\nu, \alpha_0) = \sum_{\mu, \nu} \sum_\alpha a_{\alpha\alpha}^1 \Lambda(\frac{\mu + \nu}{2}, \alpha) \Lambda(\mu, \alpha_0) \Lambda(\nu, \alpha_0) = \sum_\alpha a_{\alpha\alpha}^1 \Lambda^{(3)}(\alpha, \alpha_0) \approx [a^1_{\alpha\alpha}]_{\alpha_0}. $

Regarding the third term in \eqref{A1_g:t}, by the sparsity property we have
\begin{equation}
    \begin{split}
        &\sum_{\substack{\mu\nu\\\mu\neq\nu}}\sum_{\alpha\beta}\rho_{\alpha\beta}a^1_{\alpha\beta}\frac{\Lambda(\mu,\alpha)\Lambda(\mu,\beta)\Lambda(\nu,\alpha)\Lambda(\nu,\beta)}{\sum_\gamma\Lambda(\mu,\gamma)\Lambda(\nu,\gamma)}e^{i(E_\mu-E_\nu)t}\\&=\sum_{\substack{\mu\nu\\\mu\neq\nu}}\sum_{\alpha}\sum_{n\in N_1}\rho_{\alpha,\alpha+n}a^1_{\alpha,\alpha+n}\frac{\Lambda(\mu,\alpha)\Lambda(\mu,\alpha+n)\Lambda(\nu,\alpha)\Lambda(\nu,\alpha+n)}{\sum_\gamma\Lambda(\mu,\gamma)\Lambda(\nu,\gamma)}e^{i(E_\mu-E_\nu)t}.
    \end{split}
\end{equation}
Since off-diagonal matrix elements of $\hat{A}_1$ are present, we cannot perform an averaging procedure as above. The denominator does not cancel, so here the multiplier due to a non-Gaussian correction is of smaller order of magnitude. Moreover, $|N_1|\ll N$ and there are few contributions from the sum over $\beta$. Therefore, we can ignore terms based on a non-Gaussian correction except for the case when an observable is involved through its diagonal elements in the non-interacting basis.

Combining the results above, we get 
\begin{equation} \label{1point:C}
    \langle \Delta A_1(t)\rangle\approx\sum_{\alpha\beta}\rho_{\beta\alpha}a^1_{\alpha\beta}C_\alpha(t)C_\beta(-t)-\sum_{\alpha_0}\rho_{\alpha_0\alpha_0}[a^1_{\alpha\alpha}]_{\alpha_0}C_{\alpha_0}(t)C_{\alpha_0}(-t).
\end{equation}

\vspace{20pt}
Another observable correlation function of interest is $\langle \hat{A}_1(t)\hat{A}_2(0)\rangle$, where $\hat{A}_1$ and $\hat{A}_2$ are local observables. By the self-averaging property $\langle \hat{A}_1(t)\hat{A}_2(0)\rangle=\langle \hat{A}_1(t)\hat{A}_2(0)\rangle_V$, and Eq. \eqref{ocf} reads
\begin{equation} \label{A1A2_g}
    \langle \hat{A}_1(t)\hat{A}_2(0)\rangle =\sum_{\mu\nu\nu'}\sum_{\substack{\alpha_0\alpha\alpha'\\\beta_0\beta\beta'}}\rho_{\alpha_0\beta_0}a^1_{\alpha\beta}a^2_{\alpha'\beta'}\langle c_{\nu'}(\alpha_0)c_\mu(\beta_0)c_\mu(\alpha)c_\nu(\beta)c_\nu(\alpha')c_{\nu'}(\beta')\rangle_V e^{i(E_\mu-E_\nu)t}.
\end{equation}
Then for the time-dependent part $\langle \Delta A_1(t)A_2(0)\rangle$ we have
\begin{equation} \label{A1A2_g:t}
    \begin{split}
        \langle \Delta A_1(t)A_2(0)\rangle=&
        \sum_{\substack{\mu\nu\nu^{\prime}\\\mu\neq \nu\neq\nu^{\prime}}}\sum_{\substack{\alpha_0\alpha\alpha'\\\beta_0\beta\beta'}}\rho_{\alpha_0\beta_0}a^1_{\alpha\beta}a^2_{\alpha'\beta'}\langle c_\mu(\beta_0)c_\mu(\alpha)c_\nu(\beta)c_\nu(\alpha')c_{\nu'}(\alpha_0)c_{\nu'}(\beta')\rangle_V e^{i(E_\mu-E_\nu)t}\\&+\sum_{\substack{\mu\nu\\\mu\neq \nu}}\sum_{\substack{\alpha_0\alpha\alpha'\\\beta_0\beta\beta'}}\rho_{\alpha_0\beta_0}a^1_{\alpha\beta}a^2_{\alpha'\beta'}\langle c_\mu(\alpha_0)c_\mu(\beta_0)c_\mu(\alpha)c_\mu(\beta')c_\nu(\beta)c_\nu(\alpha')\rangle_V e^{i(E_\mu-E_\nu)t}\\
        &+\sum_{\substack{\mu\nu\\\mu\neq \nu}}\sum_{\substack{\alpha_0\alpha\alpha'\\\beta_0\beta\beta'}}\rho_{\alpha_0\beta_0}a^1_{\alpha\beta}a^2_{\alpha'\beta'}\langle c_\mu(\beta_0)c_\mu(\alpha)c_\nu(\alpha_0)c_\nu(\beta)c_\nu(\alpha')c_\nu(\beta')\rangle_V e^{i(E_\mu-E_\nu)t}.
    \end{split}
\end{equation}
We proceed by applying the same reasoning as in the case of $\langle \hat{A}_1(t)\rangle$. Analogously to the simpler case seen before, we shall denote 
\begin{equation} \label{def:Cab}
    C_{\alpha\beta}(t)=\int dE\,D(E)\Lambda(E-E_\alpha)\Lambda(E-E_\beta)e^{iEt},\hspace{20pt}C_{\alpha\beta}(-t)=\int dE\,D(E)\Lambda(E-E_\alpha)\Lambda(E-E_\beta)e^{-iEt}.
\end{equation}

By substituting the eigenstate correlation function in the first term in \eqref{A1A2_g:t} with Eq. \eqref{ecf6:3}, we get
\begin{equation} \label{2point:1st_term}
    \begin{split}
        &\sum_{\substack{\mu\nu\nu^{\prime}\\\mu\neq \nu\neq\nu^{\prime}}}\sum_{\substack{\alpha_0\alpha\alpha'\\\beta_0\beta\beta'}}\rho_{\alpha_0\beta_0}a^1_{\alpha\beta}a^2_{\alpha'\beta'}\langle c_\mu(\beta_0)c_\mu(\alpha)c_\nu(\beta)c_\nu(\alpha')c_{\nu'}(\alpha_0)c_{\nu'}(\beta')\rangle_V e^{i(E_\mu-E_\nu)t}\\
        &\approx \sum_{\alpha_0\alpha\beta}\rho_{\alpha_0\alpha}a^1_{\alpha\beta}a^2_{\beta\alpha_0}C_\alpha(t)C_\beta(-t)-\sum_{\alpha_0\alpha}\rho_{\alpha_0\alpha}a^1_{\alpha\alpha_0}[a^2_{\alpha'\alpha'}]_{\alpha_0}C_\alpha(t)C_{\alpha_0}(-t)\\
        &-\sum_{\alpha_0\beta_0}\rho_{\alpha_0\beta_0}[a^1_{\alpha\alpha}]_{\beta_0}a^2_{\beta_0\alpha_0}C_{\beta_0}(t)C_{\beta_0}(-t).
    \end{split}
\end{equation}

In order to estimate the next two terms in \eqref{A1A2_g:t} we use the eigenstate correlation function, given by Eq. \eqref{ecf6:2}. We obtain
\begin{equation} \label{2point:2nd_term}
    \begin{split}
        &\sum_{\substack{\mu\nu\\\mu\neq \nu}}\sum_{\substack{\alpha_0\alpha\alpha'\\\beta_0\beta\beta'}}\rho_{\alpha_0\beta_0}a^1_{\alpha\beta}a^2_{\alpha'\beta'}\langle c_\mu(\alpha_0)c_\mu(\beta_0)c_\mu(\alpha)c_\mu(\beta')c_\nu(\beta)c_\nu(\alpha')\rangle_V e^{i(E_\mu-E_\nu)t}\\
        &\approx\sum_{\alpha_0\beta_0\beta}\rho_{\alpha_0\beta_0}a^1_{\beta_0\beta}a^2_{\beta\alpha_0}C_{\alpha_0\beta_0}(t)C_\beta(-t)+\sum_{\alpha_0\beta_0\beta}\rho_{\alpha_0\alpha_0}a^1_{\beta_0\beta}a^2_{\beta\beta_0}C_{\alpha_0\beta_0}(t)C_\beta(-t)\\
        &+\sum_{\alpha_0\beta_0\beta}\rho_{\alpha_0\beta_0}a^1_{\alpha_0\beta}a^2_{\beta\beta_0}C_{\alpha_0\beta_0}(t)C_\beta(-t)-\sum_{\beta_0\beta}\rho_{\beta\beta_0}a^1_{\beta_0\beta}[a^2_{\alpha'\alpha'}]_{\frac{3\beta+\beta_0}{4}}C_{\beta_0\beta}(t)C_\beta(-t)\\
        &-\sum_{\alpha_0\beta_0}\rho_{\alpha_0\beta_0}a^1_{\alpha_0\beta_0}[a^2_{\alpha'\alpha'}]_{\frac{\alpha_0+3\beta_0}{4}}C_{\alpha_0\beta_0}(t)C_{\beta_0}(-t)-\sum_{\alpha_0\gamma}\rho_{\alpha_0\alpha_0}[a^1_{\alpha\alpha}]_{\frac{\alpha_0+3\gamma}{4}}[a^2_{\alpha'\alpha'}]_{\frac{\alpha_0+3\gamma}{4}}C_{\alpha_0\gamma}(t)C_\gamma(-t)\\
        &-\sum_{\alpha_0\beta_0}\rho_{\alpha_0\beta_0}[a^1_{\alpha\alpha}]_{\frac{3\alpha_0+\beta_0}{4}}a^2_{\alpha_0\beta_0}C_{\alpha_0\beta_0}(t)C_{\alpha_0}(-t)-\sum_{\alpha_0\beta_0}\rho_{\alpha_0\beta_0}[a^1_{\alpha\alpha}]_{\frac{\alpha_0+3\beta_0}{4}}a^2_{\beta_0\alpha_0}C_{\alpha_0\beta_0}(t)C_{\beta_0}(-t)
    \end{split}
\end{equation}
and
\begin{equation} \label{2point:3rd_term}
    \begin{split}
        &\sum_{\substack{\mu\nu\\\mu\neq \nu}}\sum_{\substack{\alpha_0\alpha\alpha'\\\beta_0\beta\beta'}}\rho_{\alpha_0\beta_0}a^1_{\alpha\beta}a^2_{\alpha'\beta'}\langle c_\mu(\beta_0)c_\mu(\alpha)c_\nu(\alpha_0)c_\nu(\beta)c_\nu(\alpha')c_\nu(\beta')\rangle_V e^{i(E_\mu-E_\nu)t}\\
        &\approx\sum_{\alpha_0\beta_0}\rho_{\alpha_0\beta_0}a^1_{\beta_0\alpha_0}[a^2_{\alpha'\alpha'}]_{\alpha_0}C_{\beta_0}(t)C_{\alpha_0}(-t)+\sum_{\alpha_0\beta_0\beta}\rho_{\alpha_0\beta_0}a^1_{\beta_0\beta}a^2_{\beta\alpha_0}C_{\beta_0}(t)C_{\alpha_0\beta}(-t)\\
        &+\sum_{\alpha_0\beta_0\beta}\rho_{\alpha_0\beta_0}a^1_{\beta_0\beta}a^2_{\alpha_0\beta}C_{\beta_0}(t)C_{\alpha_0\beta}(-t)-\sum_{\alpha_0\beta_0}\rho_{\alpha_0\beta_0}[a^1_{\alpha\alpha}]_{\frac{\alpha_0+3\beta_0}{4}}a^2_{\alpha_0\beta_0}C_{\beta_0}(t)C_{\alpha_0\beta_0}(-t)\\
        &-\sum_{\alpha_0}\rho_{\alpha_0\alpha_0}[a^1_{\alpha\alpha}]_{\alpha_0}[a^2_{\alpha'\alpha'}]_{\alpha_0}C_{\alpha_0}(t)C_{\alpha_0}(-t)-\sum_{\alpha_0\beta_0}\rho_{\alpha_0\beta_0}[a^1_{\alpha\alpha}]_{\frac{\alpha_0+3\beta_0}{4}}a^2_{\beta_0\alpha_0}C_{\beta_0}(t)C_{\alpha_0\beta_0}(-t).
    \end{split}
\end{equation}

Observe that for $\alpha\neq\beta$, since $\Lambda(E-E_\alpha)$ and $\Lambda(E-E_\beta)$ are centered at different points of the spectrum, it holds that $C_{\alpha\beta}(t)<C_\alpha(t)$. Therefore, the leading-order terms do not contain the quantities \eqref{def:Cab}. Further, let us consider the term $\sum_{\alpha_0}\rho_{\alpha_0\alpha_0}[a^1_{\alpha\alpha}]_{\alpha_0}[a^2_{\alpha'\alpha'}]_{\alpha_0}C_{\alpha_0}(t)C_{\alpha_0}(-t)$ in \eqref{2point:3rd_term}. The summation runs over the state space of $\hat{H}_0$ once, compared to at least twice for all other results in \eqref{2point:1st_term}-\eqref{2point:3rd_term}, so we neglect this contribution. For the time-dependent part of the two-point observable correlation function we obtain
\begin{equation} \label{2point:C}
    \langle\Delta A_1(t)A_2(0)\rangle \approx \sum_{\alpha_0\alpha\beta}\rho_{\alpha_0\alpha}a^1_{\alpha\beta}a^2_{\beta\alpha_0}C_\alpha(t)C_\beta(-t)-\sum_{\alpha_0\beta_0}\rho_{\alpha_0\beta_0}[a^1_{\alpha\alpha}]_{\beta_0}a^2_{\beta_0\alpha_0}C_{\beta_0}(t)C_{\beta_0}(-t).
\end{equation}

Due to the properties of $\Lambda(E-E_\alpha)$, mainly energies $E$ which are close to $E_\alpha$ contribute to the value of $C_\alpha(t)$. Thus we can write $D(E)=\frac{1}{\omega}$, where $\omega$ is the constant level spacing of the non-interacting Hamiltonian. Then we use the equations in \eqref{C_to_Omega} to rewrite Eq. \eqref{1point:C} and Eq.\eqref{2point:C}:
\begin{equation}
    \langle\Delta A_1(t)\rangle=\left(\langle \hat{A}_1(t)\rangle_{\hat{H}_0}-\sum_{\alpha_0}\rho_{\alpha_0\alpha_0}[a^1_{\alpha'\alpha'}]_{\alpha_0}\right)\Omega^2(t),
\end{equation}
\begin{equation}
    \langle\Delta A_1(t)A_2(0)\rangle=\left(\langle \hat{A}_1(t)\hat{A}_2(0)\rangle_{\hat{H}_0}-\sum_{\alpha_0\beta_0}\rho_{\alpha_0\beta_0}[a^1_{\alpha'\alpha'}]_{\beta_0}a^2_{\beta_0\alpha_0}\right)\Omega^2(t).
\end{equation}

We have introduced the dynamics in the non-interacting Hamiltonian, which is given by $\langle \hat{A}_1(t)\rangle_{\hat{H}_0}=\sum_{\alpha\beta}\rho_{\beta\alpha}a^1_{\alpha\beta}e^{i(E_\alpha-E_\beta)t}$ and $\langle \hat{A}_1(t)\hat{A}_2(0)\rangle_{\hat{H}_0}=\sum_{\alpha_0\alpha\beta}\rho_{\alpha_0\alpha}a^1_{\alpha\beta}a^2_{\beta\alpha_0}e^{i(E_\alpha-E_\beta)t}$.
Let us now denote by $(\cdot)_{\textrm{DE}}$ the diagonal ensemble average, and make the observation that $\langle \hat{A}_1(0)\rangle=\langle \hat{A}_1(0)\rangle_{\hat{H}_0}$. Then we have $\sum_{\alpha_0}\rho_{\alpha_0\alpha_0}[a^1_{\alpha'\alpha'}]_{\alpha_0}=(A_1)_{\textrm{DE}}$. Recall that by definition $[a^1_{\alpha'\alpha'}]_\alpha=\sum_{\alpha'}\Lambda(\alpha,\alpha')a^1_{\alpha'\alpha'}$. By the smoothness property, and slight abuse of notation, we are able to write 
\[
  (A_1)_{\textrm{DE}}=[a^1_{\alpha'\alpha'}]_{\alpha_0}\sum_{\alpha_0}\rho_{\alpha_0\alpha_0}.
\]

We thus make the replacement $[a^1_{\alpha'\alpha'}]_{\alpha_0}\to(A_1)_{\textrm{DE}}$, and obtain 
\begin{equation}
    \langle \hat{A}_1(t)\rangle=\left(\langle \hat{A}_1(t)\rangle_{\hat{H}_0}-(A_1)_{\textrm{DE}}\right)\Omega^2(t)+(A_1)_{\textrm{DE}},
\end{equation}
\begin{equation} \label{2-point-general-H_app}
\langle \hat{A}_1(t)\hat{A}_2(0)\rangle=\left(\langle \hat{A}_1(t)\hat{A}_2(0)\rangle_{\hat{H}_0}-(A_1)_{\textrm{DE}}\langle \hat{A}_2(0)\rangle\right)\Omega^2(t)+(A_1)_{\textrm{DE}}\langle \hat{A}_2(0)\rangle.
\end{equation}

To estimate the stationary part of the two-point function, we have used the initial condition $\langle \hat{A}_1(0)\hat{A}_2(0)\rangle=\langle \hat{A}_1(0)\hat{A}_2(0)\rangle_{\hat{H}_0}$.

It is important to note that the analysis above was carried out without restricting the observable correlation functions to their real-valued components, so we are able to write:
\begin{equation} \label{2-point-general-app-Im}
\textrm{Im}[\langle \hat{A}_1(t)\hat{A}_2(0)\rangle]=\textrm{Im}[\langle \hat{A}_1(t)\hat{A}_2(0)\rangle_{\hat{H}_0}]\Omega^2(t).
\end{equation}

\begin{figure} \label{fig_app_Im}
\includegraphics[width=0.48\textwidth]{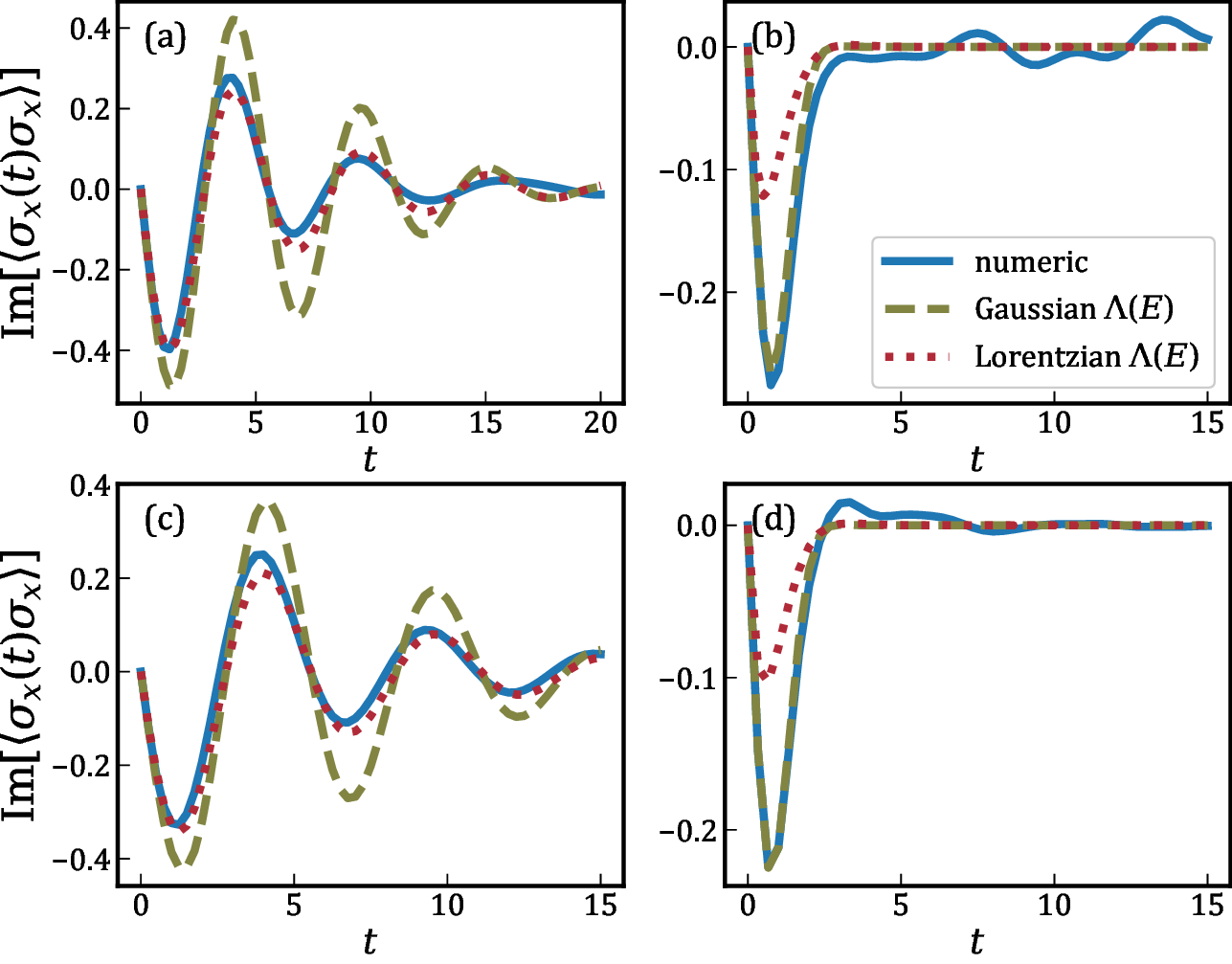}
\caption{The imaginary part of two-point observable correlation functions: (a) and (c) in weak coupling regime, (b) and (d) in strong coupling regime. Top row: the initial state is $|\Psi_0\rangle=|\phi_\alpha\rangle$,  where $\alpha=2041$. Bottom row: the initial state is random product state, numeric results are averaged over 50 realizations of the bath state. The analytic results follow Eq. \eqref{2-point-general-app-Im}.}\label{Im}
\end{figure}

With reference to Section \ref{ED} and the model discussed therein, Fig. \ref{Im} presents the time dependence of the imaginary part of autocorrelation function. In order to compare the numeric results to the analytic prediction \eqref{2-point-general-app-Im}, the integral $\Omega(t)$ is evaluated using two forms of the function $\Lambda(E)$: a Lorentzian form and a Gaussian form. The Lorentzian $\Lambda(E)$ is expected to characterize the weak coupling regime of the studied system, while the Gaussian $\Lambda(E)$ corresponds to the strong coupling regime.

\section{Two-time correlation functions}
\label{appendix:2t}
Let ($t_{1}>t_{2}$) and consider $\langle\Psi_{0}|\hat{A}_1(t_{1})\hat{A}_2(t_{2})|\Psi_{0}\rangle$. It holds that
\begin{equation}
\langle\Psi_{0}|\hat{A}_1(t_{1})\hat{A}_2(t_{2})|\Psi_{0}\rangle=\langle\Psi^{\prime}_{0}|\hat{A}_1(t_{1}-t_{2})\hat{A}_2|\Psi^{\prime}_{0}\rangle,
\end{equation}
where $|\Psi^{\prime}_{0}\rangle=e^{-i\hat{H} t_{2}}|\Psi_{0}\rangle$ is a shifted initial state. Then, following our result \eqref{2-point-general-H_app}, we get
\begin{equation}
\langle \hat{A_1}(t_{1})\hat{A}_2(t_{2})\rangle=\Big(\langle \Psi' _0| \hat{A}_1(t_{1}-t_2)\hat{A}_2|\Psi_0'\rangle_{\hat{H}_0}-(A_1)_{\textrm{DE}}\langle \hat{A_2}(t_{2})\rangle\Big)\Omega^2(|t_{1}-t_{2}|)+(A_1)_{\textrm{DE}}\langle \hat{A}_2(t_{2})\rangle.
\end{equation}

\section{Squared commutator and OTOC}
\label{appendix:otoc}
Here we present the derivation of the squared commutator $\tilde{C}(t)=\langle |[\hat{A}_1(t),\hat{A}_2(0)]|^2\rangle$. It holds that 
\begin{equation}\label{appendix:otoc_non0}  
\tilde{C}(t)=\langle\hat{A}_2(0)(\hat{A}_1(t))^2\hat{A}_2(0)\rangle+\langle\hat{A}_1(t)(\hat{A}_2(0))^2\hat{A}_1(t)\rangle-2\textrm{Re}\{\langle\hat{A}_1(t)\hat{A}_2(0)\hat{A}_1(t)\hat{A}_2(0)\rangle\}.
\end{equation}

We introduce the representation $\hat{A} = \hat{A}^0+(A)_{\textrm{DE}}\cdot\mathbbm{1}$, where the observable $\hat{A}$ is shifted to $\hat{A}^0$ which has zero DE average. Since $[\hat{A}_1(t),\hat{A_2}(0)]= [\hat{A}_1^0(t),\hat{A}_2^0(0)]$, we can rewrite \eqref{appendix:otoc_non0} in terms of shifted $\hat{A}_1$ and $\hat{A}_2$: 
\begin{equation} \label{appendix:otoc_0}    
\tilde{C}(t)=\langle\hat{A}^0_2(0)(\hat{A}_1^0(t))^2\hat{A}_2^0(0)\rangle+\langle\hat{A}_1^0(t)(\hat{A}_2^0(0))^2\hat{A}_1^0(t)\rangle-2\textrm{Re}\{\langle\hat{A}_1^0(t)\hat{A}_2^0(0)\hat{A}_1^0(t)\hat{A}_2^0(0)\rangle\}.
\end{equation}

Let us consider an observable correlation function of the type $\langle\hat{A}^0(0)\hat{B}(t)\hat{A}^0(0)\rangle$ where in general $\hat{B}$ is not shifted. By Eqs. \eqref{Heisenberg} and \eqref{ecf4} we have
\begin{equation}
            \langle\hat{A}^0(0)\hat{B}(t)\hat{A}^0(0)\rangle=\sum_{\mu\nu}\sum_{\substack{\alpha_0\alpha\alpha\\\beta_0\beta\beta}} \rho_{\alpha_0\beta_0}a^0_{\beta_0\alpha}b_{\beta\alpha'}a^0_{\beta'\alpha_0}\langle c_\mu(\alpha)c_\mu(\beta)c_\nu(\alpha')c_\nu(\beta')\rangle_Ve^{i(E_\mu-E_\nu)t}
\end{equation}
and
\begin{equation}
    \begin{split}
         \langle\Delta\hat{A}^0(0)\hat{B}(t)\hat{A}^0(0)\rangle&= \sum_{\substack{\mu\nu\\\mu\neq \nu}}\sum_{\alpha_0\beta_0\alpha\alpha'}\rho_{\alpha_0\beta_0}a^0_{\beta_0\alpha}b_{\alpha\alpha'}a^0_{\alpha'\alpha_0}\Lambda(\mu,\alpha)\Lambda(\nu,\alpha')e^{i(E_\mu-E_\nu)t}\\
        &-\sum_{\substack{\mu\nu\\\mu\neq \nu}}\sum_{\alpha_0\alpha\beta_0\beta}\rho_{\alpha_0\beta_0}a^0_{\beta_0\alpha}b_{\beta\alpha}a^0_{\beta\alpha_0}\frac{\Lambda(\mu,\alpha)\Lambda(\mu,\beta)\Lambda(\nu,\alpha)\Lambda(\nu,\beta)}{\sum_\gamma\Lambda(\mu,\gamma)\Lambda(\nu,\gamma)}e^{i(E_\mu-E_\nu)t}\\
        &-\sum_{\substack{\mu\nu\\\mu\neq \nu}}\sum_{\alpha_0\alpha\beta_0\beta}\rho_{\alpha_0\beta_0}a^0_{\beta_0\alpha}b_{\beta\beta}a^0_{\alpha\alpha_0}\frac{\Lambda(\mu,\alpha)\Lambda(\mu,\beta)\Lambda(\nu,\alpha)\Lambda(\nu,\beta)}{\sum_\gamma\Lambda(\mu,\gamma)\Lambda(\nu,\gamma)}e^{i(E_\mu-E_\nu)t}.
    \end{split}
\end{equation}

Regarding the terms based on non-Gaussian corrections, as previously discussed, the dominant contributions come from those involving diagonal elements. Therefore, we obtain
\begin{equation}
\begin{split}
    \langle\Delta\hat{A}^0(0)\hat{B}(t)\hat{A}^0(0)\rangle\approx&\sum_{\alpha_0\beta_0\alpha\alpha'}\rho_{\alpha_0\beta_0}a^0_{\beta_0\alpha}b_{\alpha\alpha'}a^0_{\alpha'\alpha_0}C_\alpha(t)C_{\alpha'}(-t)-\sum_{\alpha_0\alpha\beta_0}\rho_{\alpha_0\beta_0}a^0_{\beta_0\alpha}[b_{\beta\beta}]_{\alpha}C_\alpha(t)C_\alpha(-t)\\=&\bigg(\langle\hat{A}^0(0)\hat{B}(t)\hat{A}^0(0)\rangle_{\hat{H}_0}-(B)_{\textrm{DE}}\langle(\hat{A}^0(0))^2\rangle\bigg)\Omega^2(t).
\end{split}    
\end{equation}

Thus,
\begin{equation}
    \langle\hat{A}^0_2(0)(\hat{A}_1^0(t))^2\hat{A}_2^0(0)\rangle=\bigg(\langle\hat{A}^0_2(0)(\hat{A}_1^0(t))^2\hat{A}_2^0(0)\rangle_{\hat{H}_0}-((A^0_1)^2)_{\textrm{DE}}\langle(\hat{A}_2^0(0))^2\rangle\bigg)\Omega^2(t)+((A^0_1)^2)_{\textrm{DE}}\langle(\hat{A}_2^0(0))^2\rangle.
\end{equation}

Next we need an expression for a correlation function of the type $\langle\hat{A}^0(t)\hat{B}(0)\hat{A}^0(t)\rangle$. We can shift $\hat{B}$ so that $\langle\hat{A}^0(t)\hat{B}(0)\hat{A}^0(t)\rangle=\langle\hat{A}^0(t)\hat{B}^0(0)\hat{A}^0(t)\rangle+(B)_{\textrm{DE}}\langle(\hat{A}^0(t))^2\rangle$, where the dynamics of the latter is known. For the first term we have

\begin{equation}
    \begin{split}\langle\hat{A}^0(t)\hat{B}^0(0)\hat{A}^0(t)\rangle=&\sum_{\substack{\mu\nu\\\gamma\kappa}}\sum_{\substack{\alpha_0\alpha\alpha'\alpha''\\\beta_0\beta\beta'\beta''}}\rho_{\alpha_0\beta_0}a^0_{\alpha\beta}b^0_{\alpha'\beta'}a^0_{\alpha''\beta''}e^{i(E_\mu-E_\nu)t}e^{i(E_\gamma-E_\kappa)t}\\
    &\times\langle c_\mu(\beta_0)c_\mu(\alpha)c_\nu(\beta)c_\nu(\alpha')c_\gamma(\alpha'')c_\gamma(\beta')c_\kappa(\beta'')c_\kappa(\alpha_0)\rangle_V,
    \end{split}
\end{equation}
which depends on an eight-point eigenstate correlation function. Since the diagonal ensemble averages of the two observables are zero, the only significant contribution comes from the term without non-Gaussian corrections, 
\begin{eqnarray}
&&\langle c_\mu(\beta_0)c_\mu(\alpha)c_\nu(\beta)c_\nu(\alpha')c_\gamma(\alpha'')c_\gamma(\beta')c_\kappa(\beta'')c_\kappa(\alpha_0)\rangle_V\notag\\
&&\approx \langle c_{\mu}(\beta_{0})c_{\mu}(\alpha)\rangle_{V}\langle c_{\nu}(\beta)c_{\nu}(\alpha')\rangle_{V}
\langle c_{\gamma}(\alpha'')c_{\gamma}(\beta')\rangle_{V}\langle c_{\kappa}(\beta'')c_{\kappa}(\alpha_0)\rangle_{V}\notag\\
&&=\delta_{\alpha\beta_{0}}\Lambda(\mu,\alpha)\delta_{\beta\alpha'}\Lambda(\nu,\beta)\delta_{\beta'\alpha''}\Lambda(\gamma,\beta')\delta_{\beta''\alpha_0}\Lambda(\kappa,\beta'').
\end{eqnarray}

Then we have
\begin{equation}
    \begin{split}\langle\Delta\hat{A}^0(t)\hat{B}^0(0)\hat{A}^0(t)\rangle=&\sum_{\substack{\alpha_0\alpha\\\beta\beta'}}\rho_{\alpha_0\alpha}a^0_{\alpha\beta}b^0_{\beta\beta'}a^0_{\beta'\alpha_0}e^{i(E_\mu-E_\nu)t}e^{i(E_\gamma-E_\kappa)t} C_\alpha(t)C_\beta(-t)C_{\beta'}(t)C_{\alpha_0}(-t),
    \end{split}
\end{equation}
which leads to
\begin{equation}
    \langle\hat{A}^0(t)\hat{B}^0(0)\hat{A}^0(t)\rangle=\langle\hat{A}^0(t)\hat{B}^0(0)\hat{A}^0(t)\rangle_{\hat{H}_0}\Omega^4(t).
\end{equation}

The above result and Eq. \eqref{2-point-general-H_app} imply that the second term in Eq. \eqref{appendix:otoc_0} has the form
\begin{equation}
\begin{split}
    \langle\hat{A}_1^0(t)(\hat{A}_2^0(0))^2\hat{A}_1^0(t)\rangle &= \langle\hat{A}_1^0(t)\big((\hat{A}_2^0(0))^2\big)\hat{A}_1^0(t)\rangle_{\hat{H}_0}\Omega^4(t)+\bigg(\langle(\hat{A}_1^0)^2(t)\rangle_{\hat{H}_0}-\big((A_1^0)^2\big)_{\textrm{DE}}\bigg)\big((A_2^0)^2\big)_{\textrm{DE}} \Omega^2(t)\\&+\big((A_1^0)^2\big)_{\textrm{DE}}\big((A_2^0)^2\big)_{\textrm{DE}} .
\end{split}
\end{equation}

The third term in Eq. \eqref{appendix:otoc_0} is a particular case of the observable correlation given by $\langle \hat{A}^0(t)\hat{B}^0(0)\hat{V}^0(t)\hat{W}^0(0)\rangle$, where the diagonal ensemble average of each observable is zero. Arguing as above, we have
\begin{equation}
    \begin{split}
        \langle \hat{A}^0(t)\hat{B}^0(0)\hat{V}^0(t)\hat{W}^0(0)\rangle=\sum_{\mu\nu\gamma\kappa}\sum_{\substack{\alpha_0\alpha\alpha'\alpha''\alpha'''\\\beta_0\beta\beta'\beta''}}\rho_{\alpha_0\beta_0}a^0_{\alpha\beta}b^0_{\alpha'\beta'}v^0_{\alpha''\beta''}w^0_{\alpha'''\alpha_0}e^{i(E_\mu-E_\nu)t}e^{i(E_\gamma-E_\kappa)t}\\
        \times\langle c_\mu(\beta_0)c_\mu(\alpha)c_\nu(\beta)c_\nu(\alpha')c_\gamma(\beta')c_\gamma(\alpha'')c_\kappa(\beta'')c_\kappa(\alpha''')\rangle_V
    \end{split}
\end{equation}
and
\begin{equation}
        \langle \Delta\hat{A}^0(t)\hat{B}^0(0)\hat{V}^0(t)\hat{W}^0(0)\rangle=\sum_{\substack{\alpha_0\alpha\\\beta\beta'\beta''}}\rho_{\alpha_0\alpha}a^0_{\alpha\beta}b^0_{\beta\beta'}v^0_{\beta'\beta''}w^0_{\beta''\alpha_0}C_\alpha(t)C_\beta(-t)C_{\beta'}(t)C_{\beta''}(-t),
\end{equation}
therefore,
\begin{equation}
\langle \hat{A}^0(t)\hat{B}^0(0)\hat{V}^0(t)\hat{W}^0(0)\rangle=\langle \hat{A}^0(t)\hat{B}^0(0)\hat{V}^0(t)\hat{W}^0(0)\rangle_{H_0}\Omega^4(t).
\end{equation}

The result for the squared commutator reads 
\begin{equation}
\begin{split}
    \tilde{C}(t) = &\bigg(-2\langle \hat{A}^0_1(t)\hat{A}^0_2(0)\hat{A}^0_1(t)\hat{A}^0_2(0)\rangle_{\hat{H}_0}+\langle\hat{A}_1^0(t)\big((\hat{A}_2^0(0))^2\big)\hat{A}_1^0(t)\rangle_{\hat{H}_0}\bigg)\Omega^4(t)\\
    +&\bigg(\langle\hat{A}^0_2(0)(\hat{A}_1^0(t))^2\hat{A}_2^0(0)\rangle_{\hat{H}_0}+\langle(\hat{A}_1^0(t))^2\rangle_{\hat{H}_0}\big((A_2^0)^2\big)_{\textrm{DE}} \\-&((A^0_1)^2)_{\textrm{DE}}\langle(\hat{A}_2^0(0))^2\rangle-\big((A_1^0)^2\big)_{\textrm{DE}}\big((A_2^0)^2\big)_{\textrm{DE}}\bigg)\Omega^2(t)\\+&((A^0_1)^2)_{\textrm{DE}}\langle(\hat{A}_2^0(0))^2\rangle+\big((A_1^0)^2\big)_{\textrm{DE}}\big((A_2^0)^2\big)_{\textrm{DE}}.
\end{split}    
\end{equation}

\end{widetext}

\section{Random product state as the initial state} \label{appendix:random_initial_state}
In the following, we present some numerical results for the spin chain model introduced in Section \ref{ED}. These are based on initial states chosen as random product states, i.e.
\begin{equation}
    |\Psi_0\rangle=\bigotimes_{i=1}^N (\cos\theta_i\,\mid\uparrow_i\rangle+\sin\theta_i\,\mid\downarrow_i\rangle),
\end{equation}
where $N$ is the length of the chain, and $\theta_i$ are independent random variables uniformly distributed on $[0,2\pi]$, each associated with a site of the spin chain. In order to recover the typical behavior of the observable correlation functions, we average the numeric results over many realizations of the bath state, such that $\langle\Psi_0|\hat{H}_0|\Psi_0\rangle$ is close to the middle of the spectrum of $\hat{H}_0$.

\begin{figure} 
\includegraphics[width=0.48\textwidth]{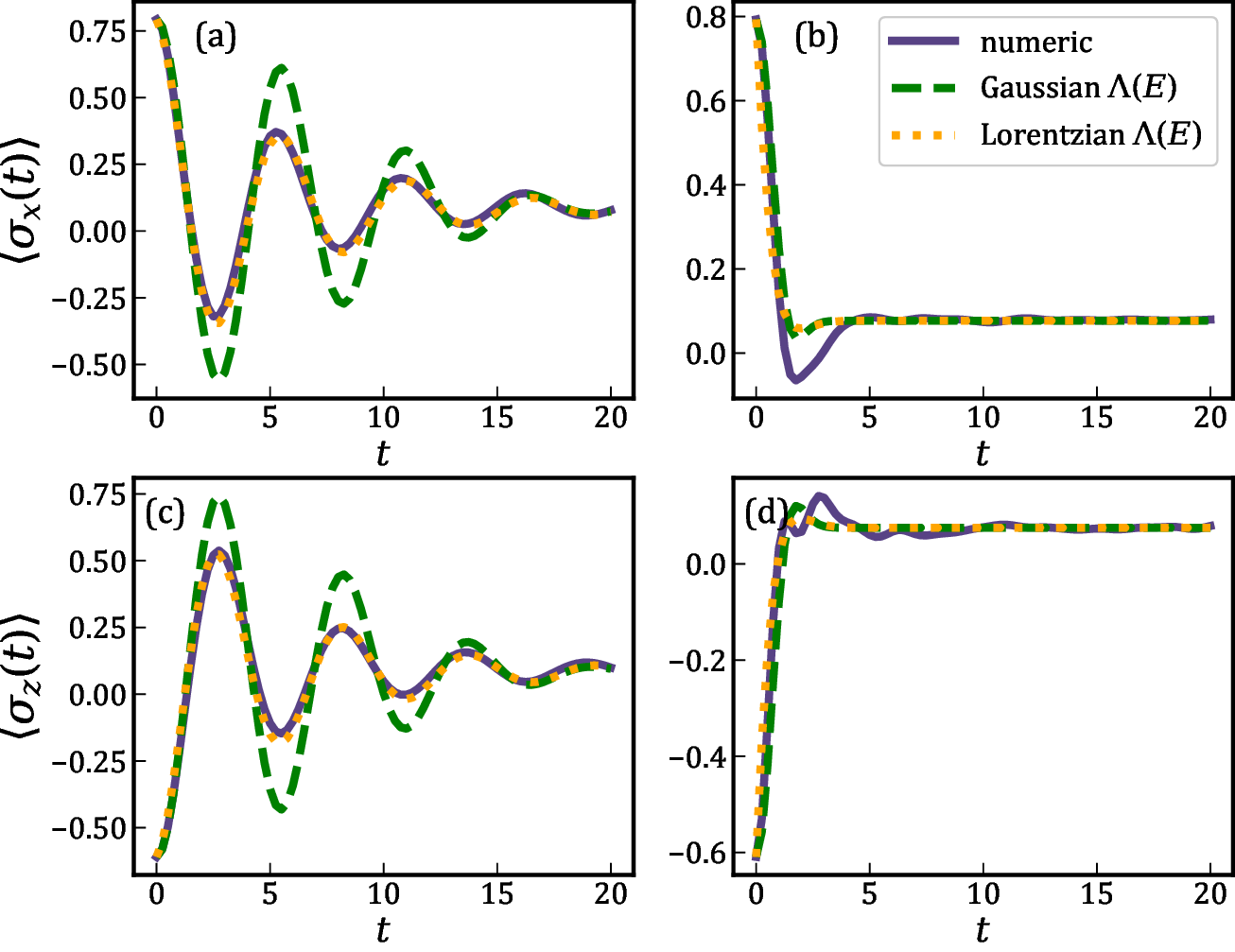}
\caption{One-point observable correlation functions: (a) and (c) in weak coupling regime, $J_x^{\rm i}=0.1$, (b) and (d) in strong coupling regime, $J_x^{\rm i}=0.8$. The numeric results are averages over 50 realizations of the bath initial state. The analytic results are given by \eqref{1-point-weak} and \eqref{1-point-strong}. The system consists of 12 spins and the other parameters are set to $B_z^{\rm s}=B_z^{\rm s}=0.4$, $B_x^{\rm b} = 0.3$,  $J_x^{\rm b}=0.7$, $J_z^{\rm i}=0.2$, $r_1=5$, $r_2=10$. We work with $\Gamma=0.087$ and $K=0.005$ for weak coupling, and  $\Gamma=0.79$, $K=0.31$ for strong coupling.}
\label{apF1}
\end{figure}

\begin{figure} [H]
\includegraphics[width=0.48\textwidth]{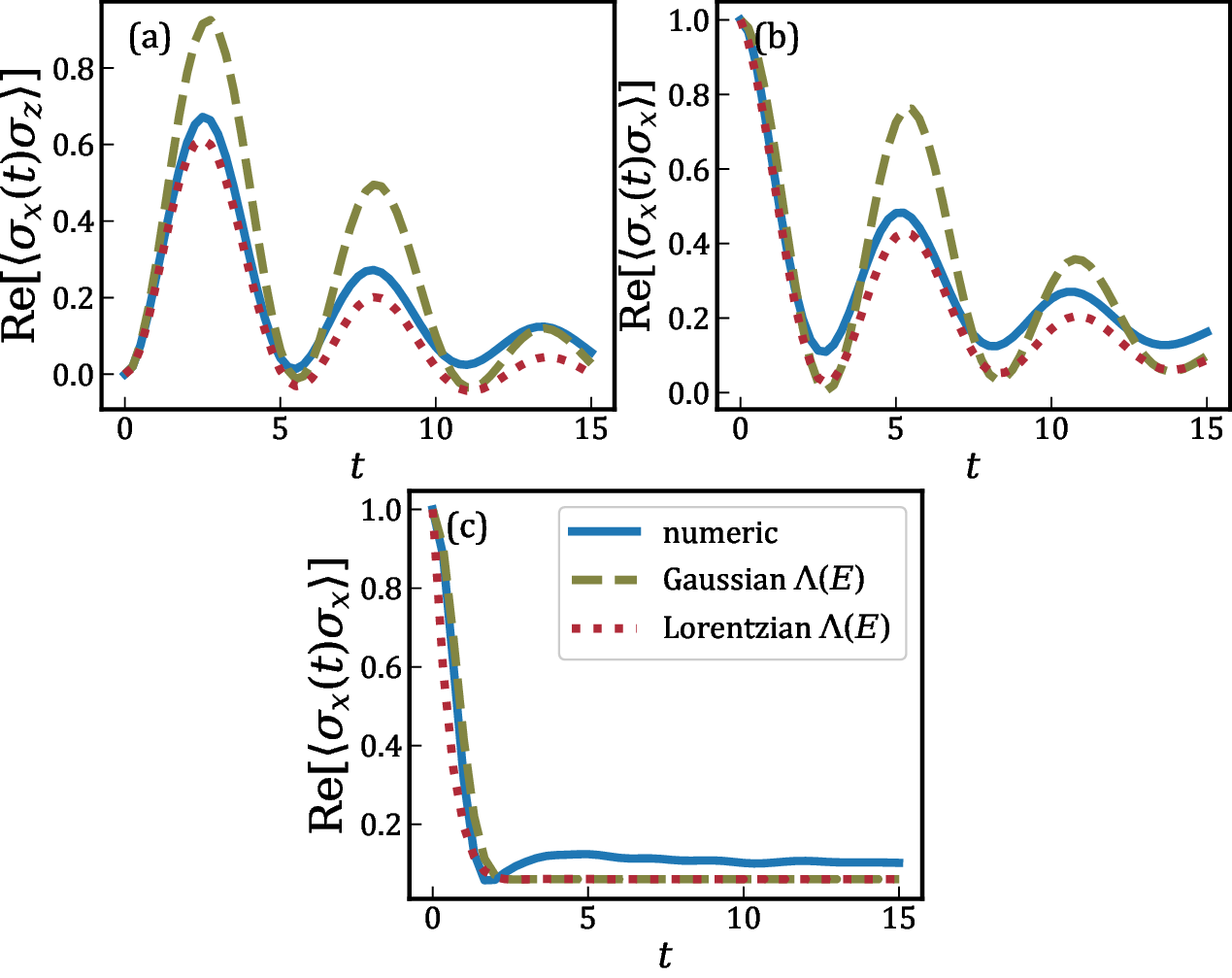}
\caption{The real part of two-point observable correlation functions: (a) and in weak coupling regime, (b) and (c) in strong coupling regime. The analytic results are given by \eqref{2-point-weak} and \eqref{2-point-strong}.}
\label{apF2}
\end{figure}

\begin{figure} [H]
\includegraphics[width=0.48\textwidth]{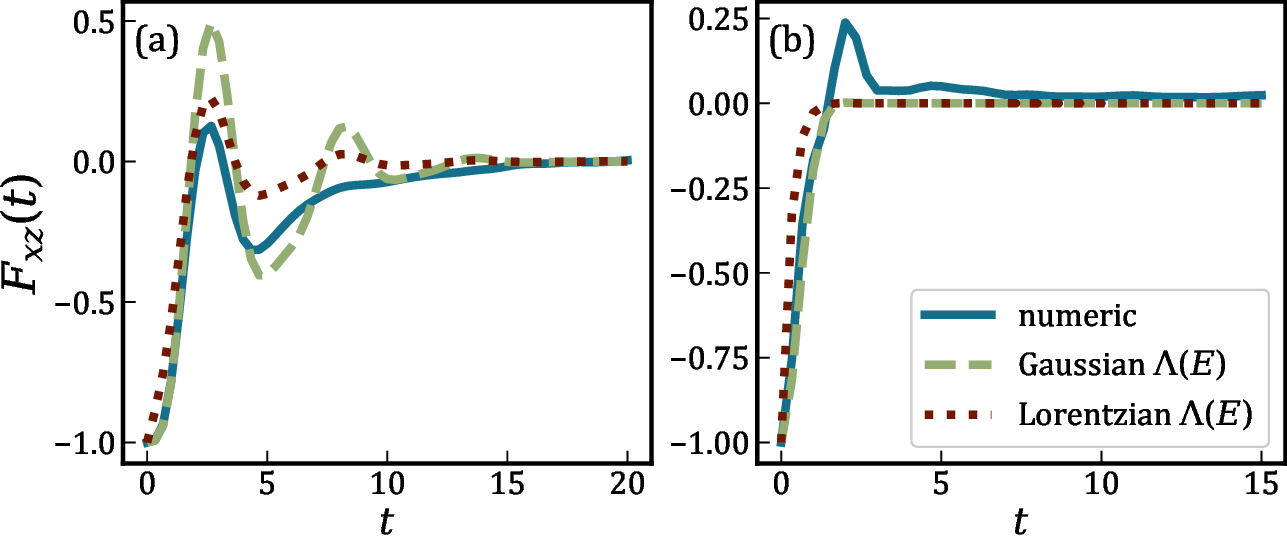}
\caption{Out-of-time-ordered correlator $F_{xz}(t)=\langle\sigma_x(t)\sigma_z\sigma_x(t)\sigma_z\rangle$ in weak coupling regime (a), and in strong coupling regime (b). The analytic results are given by \eqref{4-point-weak} and \eqref{4-point-strong}.}
\label{apF3}
\end{figure}

In both weak and strong coupling regimes, we study the time dependence of the expectation values, Fig. \ref{apF1}, the two-point observable correlation functions, Fig. \ref{apF2}, and the OTOC, Fig. \ref{apF3}. The numerical data aligns well with the analytical predictions, with most of the fluctuations averaged out.


Outside weak coupling regime, we consistently see a fluctuation in the early stages of the evolution, which does not average out, see Fig. \ref{apF1} (b) and (d), Fig. \ref{apF2} (c), and Fig. \ref{apF3} (b). A simple numerical check shows that varying the positions $r_1$ and $r_2$ affects the amplitude of such fluctuations. As a result, around the predicted relaxation time, the dynamics form a band, rather than a single trajectory, as indicated by the analytical results. This effect is especially pronounced when only one bath spin is coupled to the subsystem. Observable correlations of all studied orders exhibit this feature. We conclude that the chain geometry, together with possible edge effects, influences the characteristic dynamics observed during the early stages of the evolution.
\vspace{15pt}

\end{document}